\documentclass[12pt]{article} 

\usepackage{tocloft}
\usepackage{graphics}
\usepackage{graphicx}
\usepackage{booktabs}
\usepackage[caption=false]{subfig}
\usepackage{amsmath}
\usepackage{amsfonts}
\usepackage{amssymb}
\usepackage{verbatim}
\usepackage[margin=1in]{geometry}
\usepackage{xcolor}
\usepackage{hyperref}
\hypersetup{
    colorlinks,
    citecolor=blue!50!black,
    filecolor=black,
    linkcolor=red!50!black,
    urlcolor=red!80!black
}
\usepackage{amsthm}
\usepackage{enumitem}
\usepackage{bbm}
\theoremstyle{definition}
\newtheorem{definition}{Definition}
\newtheorem{example}{Example}

\counterwithin{subexample}{example}
\graphicspath{
{figures/}
}
\usepackage[round]{natbib}
\usepackage{array}
\usepackage{tikz}

\newcommand{\semantics}[1]{[\![\textnormal{$ #1 $\/}]\!]}

\def\citepos#1{\citeauthor{#1}'s (\citeyear{#1})}

\title{A Communication-First Account of Explanation}
\date{Stanford University}
\author{Jacqueline Harding, Tobias Gerstenberg, Thomas Icard}

\begin{document}

\maketitle

\begin{abstract}
This paper develops a formal account of causal explanation, grounded in a theory of conversational pragmatics, and inspired by the interventionist idea that explanation is about asking and answering what-if-things-had-been-different questions. We illustrate the fruitfulness of the account, relative to previous accounts, by showing that widely recognised ``explanatory virtues'' emerge naturally, as do subtle empirical patterns concerning the impact of norms on causal judgments. This shows the value of a ``communication-first'' approach to explanation: getting clear on explanation's communicative dimension is an important prerequisite for philosophical work on explanation. The result is a simple but powerful framework for incorporating insights from the cognitive sciences into philosophical work on explanation, which will be useful for philosophers or cognitive scientists interested in explanation.
\end{abstract}

\section{Introduction}\label{section: introduction}

The philosophical problem of explanation is usually taken to be a matter of locating the right \emph{explanatory relations} in the world, with human interests and capacities offered a marginal role. Although philosophers of explanation will agree that some explanatory virtues are better accounted for in terms of `pragmatic' factors (those involving background knowledge or downstream interests), they take this to show that these virtues are peripheral to the core of the explanatory project. \citet[p.7]{strevens2008DepthAccountScientific} gives clear voice to this perspective: ``To discern the nature of explanation in the ontological sense, you must acquire the ability to see past the communicative conventions and strategies of scientists to the explanatory facts themselves. A philosopher of explanation will therefore occasionally discuss communicative conventions, just as an astronomer might study atmospheric distortion so as to more clearly see the stars''.

It is tempting, then, to think of the philosophy of explanation and the psychology of explanation as separate subjects. While the philosophical literature has exerted considerable influence on the way psychologists investigate explanation \citep{hilton1990ConversationalProcessesCausal, lombrozo2006StructureFunctionExplanations,Keil2006,goddu2024development}, philosophers have been more reluctant to incorporate insights from psychology and cognitive science into their accounts. Even interventionists, who motivate their account by gesturing to the downstream usefulness of causal information, eschew pragmatic factors in the account's actual development \citep{woodward2003MakingThingsHappen, HitchcockWoodward1}.

Here's the question, then, which motivates our paper: what would happen if we `put communication first' \citep{potochnik2016ScientificExplanationPutting} in developing an account of explanation? Building on recent advances in cognitive science and formal pragmatics \citep{frank2012predicting, goodman2016pragmati, degen2023rational,sumers2023reconciling,beller2024language}, we direct attention to the question of what a speaker is doing when she offers an explanation to a listener. Our main claim is that several hallmarks of explanation -- including key features that were explicitly built in to previous formalisations -- simply (and softly) emerge from the dynamics of conversation together with the minimal assumption that ``why'' and ``because'' communicate relations of (typically causal) dependence.

Many philosophers have stressed the relevance of conversational and cognitive considerations in characterising explanation \citep{bromberger1965ApproachExplanation,bromberger1984PragmaticScientificExplanation,vanfraassen1977PragmaticsExplanationb,vanfraassen1980ScientificImage,achinstein1983NatureExplanation,ylikoski2010DissectingExplanatoryPower,deregt2005ContextualApproachScientifica,deRegt2017, potochnik2016ScientificExplanationPutting, potochnik2017IdealizationAimsScience}, but these proposals have largely remained programmatic and informal.\footnote{Attempts to treat explanation formally have tended to ignore pragmatic factors. The first formal theories of explanation involved sets of sentences closed under logical deduction (that is, logical ``theories''; see \citet{hempel1948StudiesLogicExplanation}), statistical models and probability distributions 
\citep{DouvenAbduction,hempel1965AspectsScientificExplanation,salmon1971StatisticalExplanationStatistical}, and even physical models of spacetime \citep{dowe,salmon1998CausalityExplanation}. Some of these accounts are given an ontic gloss (e.g. \citealt{salmon1998CausalityExplanation}), in that the components are intended to refer to objective features of the world; others are given a more epistemic gloss  (e.g. \citealt{hempel1948StudiesLogicExplanation}), in that they are relative to a body of evidence.} As \cite{woodward2021ScientificExplanation} have pointed out \citep[see also][]{WoodwardHuman}, most prominent philosophical accounts of explanation that stress the importance of psychological facts and interests nevertheless engage very little with work in the cognitive sciences on how people in fact produce and interpret explanations. By contrast, our formal account is developed using tools from contemporary cognitive science; this suggests a promising future avenue for incorporating psychological work into the philosophy of explanation.

\section{Preliminaries: Context and Causation}\label{section: explanation and causation}

The starting point for `pragmatic' theories of explanation is a simple observation: explanations are acts of communication, paradigmatically answers to ``why?'' questions \citep{hempel1948StudiesLogicExplanation, bromberger1965ApproachExplanation}. This suggests that to analyse explanation, one must account for the context of communication.

\subsection{Communicative Context}\label{section: communicative context}

Communicative context affects what is being explained. Asking ``Why $\mathsf{FACT}$?'' (where $\mathsf{FACT}$ is an \textit{explanandum}, a proposition to be explained) is typically understood relative to a set of relevant alternatives to $\mathsf{FACT}$, which varies with context (\citealt{vanfraassen1980ScientificImage}; see also \citealt[Section 5.12] {woodward2003MakingThingsHappen} and \citealt{ylikoski2007IdeaContrastiveExplanandum}). The question ``Why did Adam eat the apple?'' can -- depending on whether stress is placed on ``Adam'' or ``apple'' \citep{dretske} -- be a request for information about why Adam (rather than someone else) ate the apple, or about why Adam ate the apple (rather than something else) \citep[p.127]{vanfraassen1980ScientificImage}.

Most philosophers will grant this. Pragmatic theories, though, take this idea further: to judge whether a candidate explanation is successful, one must know what prompted the need for the explanation in the first place, and what it would take to satisfy that need. Consider a variant of an example from \cite{gardenfors1980PragmaticApproachExplanations}.

\begin{example}[Holiday Tan]\label{example: holiday tan}
Alice sees her friend Bob after returning from a work trip to Rome. She is noticeably more tanned than when Bob last saw her; he asks, `why are you so tanned?'. She replies, `I was just in Rome!'.
\end{example}
We can make two observations here. First, given what Bob already knows, he can -- upon hearing Alice's answer -- reasonably infer that it was sunny when she was in Rome, that she spent time outside while she was there, and so on. What makes it seem like a good explanation is, at least in part, that Bob is able to draw all of these conclusions from Alice's answer. Second, if Bob also happens to know that it was raining in Rome all week, then it is not as good of an explanation. It may also be a poor explanation if Bob already knows that Alice was in Rome. In such contexts, the explanatory question evidently persists even after Alice's statement.

Both observations suggest that communicative context is relevant to assessing the explanation given. When successful, Alice's answer to Bob's question \textit{demystifies} a fact that was initially surprising, or worthy of explanation. Given what Bob already knows, he is able to trace the sequence of events that led to Alice's tan; rather than being peripheral to explanation, contextual facts (such as Bob's background knowledge and downstream interests) seem crucial in accounting for why we have a practice of giving and receiving explanations at all.

Let $\mathcal{K}$ be the \emph{epistemic state} of the receiver of the explanation (Bob, in the example above).\footnote{To standardise terminology, we refer to the receiver as a `listener' and the explainer as a `speaker'. Of course, not all explanations are spoken.} The discussion above suggests that explanation can only be characterised and evaluated relative to this epistemic state $\mathcal{K}$. \citet{gardenfors1980PragmaticApproachExplanations,Gardenford1988,gardenfors1990EpistemicAnalysisExplanations} develops this; he assumes that facts demanding explanation are those which are surprising to an agent, in the sense that, prior to learning them, the agent assigned them relatively low probability.\footnote{This idea is developed more recently by \citet{Chandra2024}.} Explaining why $\mathsf{FACT}$ is a matter of specifying information that would have -- in the agent's epistemic state prior to learning $\mathsf{FACT}$ -- rendered it less surprising. (In the example above, the statement that Alice was in Rome counts as an explanation because it renders her tan less surprising to Bob, given his epistemic state.)

\subsection{Causation} \label{section:interventions}

Pragmatic approaches analyse explanation by emphasising its similarity to other acts of communication. A common worry, then, is that they cannot distinguish explanation from other informative acts of communication. To borrow an example from \citet[p.34]{salmon1971StatisticalExplanationStatistical} discussed by \citet[Sect. 4.2.]{woodward2003MakingThingsHappen}, learning that a cisgender man takes birth control pills may -- given suitable background assumptions about one's epistemic state\footnote{These assumptions will include (a) ignorance about whether cisgender men can get pregnant; (b) knowledge that taking birth control pills prevents pregnancy.} -- lower one's surprise at his failure to become pregnant (i.e. satisfy \citepos{gardenfors1980PragmaticApproachExplanations} criterion), but it does not \textit{explain} this failure. This intuition is reflected in the distinctive grammar of answers to ``why?'' questions; explanations can be prefixed with ``because'' in a way that other acts of communication cannot.

To account for this, a common view is that explanation fundamentally involves relations of asymmetric dependence, chiefly causal dependence \citep{HitchcockWoodward1, woodward2003MakingThingsHappen, strevens2008DepthAccountScientific}. This view is nicely summarised by \citet[p. 162]{Salmon1977},  ``To give [\dots] explanations is to show how events [\dots] fit into the causal network of the world''. The thought is that, e.g., telling someone that you took a walk in the park today (a description of \emph{what} happened) is not typically an explanation, unless it helps address a causal-explanatory question (e.g., \emph{why} your shoes were muddy, what 
\emph{caused} them to be muddy). As Salmon and Woodward observe, the issue in the example above is that the man's taking birth control pills does not play a causal role in his failure to become pregnant. Indeed, in \autoref{example: holiday tan}, Alice's response seems like an explanation only if it provides information about what \textit{caused} her tan. It may be helpful for some other purpose; it would just not count as an explanation.

Contemporary formal approaches to causation are centred around the notion of \emph{causal intervention} (see \citealt{Spirtes1993,Pearl1995,Hitchcock2001,woodward2003MakingThingsHappen}, among many others). An intervention on a system is roughly an ideal, exogenous manipulation of some component of the system, which leaves all other components unchanged. This idea has been made more precise with the help of various mathematical models, among the most general of which is the \emph{structural causal model} (see \citealt{peters2017elements,bareinboim:etal20}):
\begin{definition}[Structural Causal Model]\label{definition: scm}
A structural causal model (SCM) $\mathcal{M}$ is a $4$-tuple
\begin{equation*}
    (\mathbf{U}, \mathbf{V}, \mathbf{f_{V}},P(\mathbf{U}))
\end{equation*}
where:
\begin{itemize}
    \item $\mathbf{U}$ is a set of \emph{exogenous variables}, with possible values $\mathsf{Val}(U)$ for each $U \in \mathbf{U}$;
    \item $\mathbf{V}$ is a set of \emph{endogenous variables}, with possible values $\mathsf{Val}(X)$ for each $X \in \mathbf{V}$;\footnote{We are assuming, without loss, that $\mathsf{Val}(X) \cap \mathsf{Val}(Y) = \emptyset$ whenever $X \neq Y$.
Note also that $\mathsf{Val}(\mathbf{X}) = \bigcup_{X \in \mathbf{X}}\mathsf{Val}(X)$.}
    \item $\mathbf{f_{V}}$ is a set of \emph{structural functions}, where $f_X:\mathsf{Val}(\mathbf{V}\cup\mathbf{U}) \rightarrow \mathsf{Val}(X)$ for each endogenous variable $X \in \mathbf{V}$;
    \item $P(\mathbf{U})$ is a probability distribution over $\mathsf{Val}(\mathbf{U})$.
\end{itemize}
\end{definition}

\noindent The notion of \emph{intervention} on an SCM is captured by mechanism replacement: intervening to set variable $X$ to value $x$ is a matter of replacing the structural function $f_X$ with the constant function sending all arguments to $x$.

SCMs have risen to prominence not only in philosophy, but also in psychology and cognitive science, where a number of researchers have suggested that intuitive causal cognition employs representations, and operations over them, that resemble causal models with interventions \citep{lombrozo2006StructureFunctionExplanations,lagnado2013causal,rottman2016do-peopl,waldmann2017the-oxfo,goddu2024development}.

Consider the simple scenario in \autoref{example: holiday tan}. We model this with three endogenous variables $A,B,C$, each binary valued, so that $\mathsf{Val}(X) = \{0,1\}$ for $X = A,B,C$, and three binary exogenous variables $U_A,U_B,U_C$. $E$ represents whether Alice is tanned ($E=1$ when she is, $0$ when not); $A$ represents whether Alice is in Rome; $B$ represents whether it is sunny in Rome; and $C$ represents some other cause of her tan (e.g. whether she went to a tanning salon). The structural functions have $E$ true when either $C$ is true, or \emph{both} $A$ and $B$ are true.

\begin{figure}[ht]\centering
\begin{tikzpicture}
    \tikzset{node style/.style={circle, draw, minimum size=.8cm, fill opacity=1.0}}
    
    \node at (-3.5,0) {};

    \node[circle,draw=black,thick,dotted] (UA) at (0,1.5) {\footnotesize{}$U_A$};
    \node[circle,draw=black,thick,dotted] (UC) at (3,1.5) {\footnotesize{}$U_C$};
    \node[circle,draw=black,thick,dotted] (UB) at (1.5,1.5) {\footnotesize{}$U_B$};
    
    \node[node style] (A) at (0,0) {$A$};
    \node[node style] (B) at (1.5,0) {$B$};
    \node[node style] (C) at (3,0) {$C$}; 
    \node[node style] (E) at (1.5,-1.5) {$E$};
    
    \draw[->,thick] (A) -- (E);
    \draw[->,thick] (B) -- (E);
    \draw[->,thick] (C) -- (E); 
    \draw[->,dotted,thick] (UA) -- (A);
    \draw[->,dotted,thick] (UB) -- (B);
    \draw[->,dotted,thick] (UC) -- (C);
    \node at (5,-1) {$f_E(a,b,c) = \max(c, \min(a,b))$};
    \node at (-.4,.8) {$f_A$};
    \node at (1.1,.8) {$f_B$};
    \node at (2.6,.8) {$f_C$};
\end{tikzpicture}
    \caption{Causal Model $\mathcal{M}_{(A\land B)\lor C}$.}
    \label{fig:holidaytanstructure}
\end{figure}
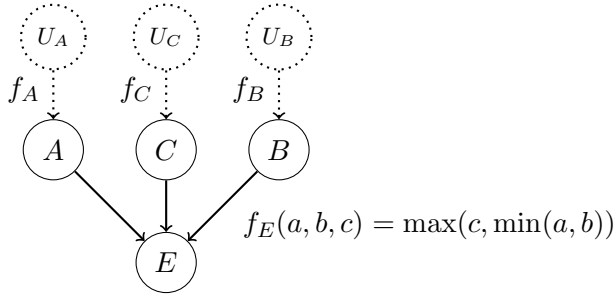

Meanwhile, $A$, $B$, and $C$ are all determined exogenously, taking on the value of their respective exogenous variables ($U_A$, $U_B$, and $U_C$). That is, $f_A$, $f_B$, and $f_C$ are all the identity function.  See \autoref{fig:holidaytanstructure}, where arrows represent functional relationships. (To fully specify the model $\mathcal{M}_{(A\land B)\lor C}$, we would also need to define a probability distribution $P(U_A,U_B,U_C)$, which in the simplest cases would factor as a product $P(U_A)\cdot P(U_B)\cdot P(U_C)$.)

How might we use SCMs to model explanation? Our starting point (following \citealt{woodward2003MakingThingsHappen, halpern2005causesExplanationsCause}) is to represent an explanandum $\mathsf{FACT}$ as a proposition $\mathbf{X}=\mathbf{x}$ (the proposition that variables $\mathbf{X}$ take value $\mathbf{x}$). In the example above, then, the explanandum is $E=1$ (the proposition that Alice is tanned).\footnote{For convenience, we will sometimes refer to propositions $\mathbf{X}=\mathbf{x}$ as `events'. This should not be taken to imply that the only explananda/explanantia our model can accommodate are singular events; as the example above illustrates, it's clear that propositions $\mathbf{X}=\mathbf{x}$ can represent regularities, a point which is recognised by interventionists.}

In developing our model, we also help ourselves to a notion of ``actual causation''. An ``actual cause'' of some token event $E=1$ is an event $A=1$ that causally contributed to $E=1$ \citep{halpern2005CausesExplanationsStructuralModela}. A common way to think about this causal contribution is counterfactual, employing a `but-for' test: if $E=1$ would not have occurred \emph{but for} $A=1$, then we say that $A=1$ was an actual cause of $E=1$. In \autoref{example: holiday tan}, Alice's being in Rome was an actual cause of her tan -- had she not been in Rome, she would not have tanned. 

While this simple but-for analysis is widely applicable, we often want to identify some factor as an actual cause even when it fails the but-for test. Suppose, for example, that Alice not only went to sunny Rome ($A=1$ and $B=1$), but she also went to a tanning salon ($C=1$). We might want to identify the tanning salon as a cause of her tan even though, had she not gone, the Roman sun still would have tanned her. Such cases of \emph{overdetermination} -- and  many other puzzle cases for the but-for test -- have motivated a variety of analyses of actual cause (see, e.g., \citealt{Hitchcock2001,woodward2003MakingThingsHappen,halpern2016causality, Gallow2021}, among many others).

The rough idea behind these and other counterfactual analyses is that $C=1$ should be a but-for cause under some contingency in which we hold the values of some variables (outside a ``causal pathway'' from $C=1$ to $E=1$) fixed to possibly non-actual values. In \autoref{example: holiday tan}, note that $C=1$ is a but-for cause of $E=1$ if, for instance, we imagine what would have happened had Alice not been in Rome (that is, assume $A=0$). Different proposals make this intuition precise in subtly different ways. Such nuances will not matter for our purposes; we only need to assume that we have fixed some ``egalitarian'' analysis of what it is to be a causally contributing factor \citep{bebb2024CausalSelectionEgalitarianism}. In particular, our proposal is compatible with any account of actual causation which treats both events as actual causes in cases of overdetermination. In the case above, for example, we assume that $A=1$, $B=1$ and $C=1$ \textit{all} count as actual causes of $E=1$.

How exactly do causal interventions fit into a broader theory of explanation? Proponents of interventionist approaches to explanation agree with something like the following picture, articulated by \cite[p. 21]{HitchcockWoodward1}: ``To explain why some phenomenon occurs is to [provide] the resources for answering a variety of what-if-things-had-been-different questions: how would the outcome have differed if the initial conditions had been changed in various ways?'', where `changed' is to be understood as, `changed by ideal intervention'.

While this is a rather different conception from the approaches to explanation that came before, it leaves open lots of important details. Which what-if-things-had-been-different questions are most important? What kinds of ``resources'' for answering such questions are suitable, and how are they to be provided?

We propose that a natural approach to answering these questions pays attention to the communicative context. This approach combines the interventionist idea that explanations communicate patterns of dependence with explicit modelling of more `pragmatic' factors, such as the epistemic state of the listener.

\section{A Communication-First Account}

As we see it, the clearest development to date of the hybrid approach suggested above comes from \cite{halpern2005causesExplanationsCause} (though the authors don't present it as such). Although our account diverges significantly from theirs, its starting point is similar.\footnote{Halpern and Pearl's (HP) analysis -- which has influenced not only philosophy, but also cognitive science and computer science -- defines an explanation of a proposition $\mathsf{FACT}$ as a proposition $\mathbf{X}=\mathbf{x}$ such that: \textbf{(EX1)} $\mathsf{FACT}$ is true at all models $(\mathcal{M}, \mathbf{u})\in \mathcal{K}$, \textbf{(EX2)} $\mathbf{X} = \mathbf{x}$ is an actual cause of $\mathsf{FACT}$ for all $(\mathcal{M}, \mathbf{u})\in \mathcal{K}$ in which $\mathbf{X} = \mathbf{x}$ is true, \textbf{(EX3)} No proper subset of $\mathbf{X}$ satisfies EX2 and \textbf{(EX4)} There exists $(\mathcal{M}, \mathbf{u}) \in \mathcal{K}$ in which $\mathbf{X} \neq \mathbf{x}$. Although we accept EX1, we deny EX2, EX3 and EX4, as discussed in \autoref{section: explaining explanation}.} Following \citet{halpern2005causesExplanationsCause}, we suppose that there is a listener whose uncertainty is represented by a probability distribution $\mathsf{Prior}$ on a set $\mathcal{K}$ of pairs $(\mathcal{M}, \mathbf{u})$, where $\mathcal{M}$ is a causal model and $\mathbf{u}$ is a ``context'' that specifies values for all the exogenous variables.\footnote{In general, $\mathsf{Prior}$ is a distribution on a $\sigma$-algebra over $\mathcal{K}$. It is supposed for simplicity that all models in $\mathcal{K}$ have the same variables, so that uncertainty is only over the structural relationships and the values of (both exogenous and endogenous) variables.} Intuitively, these are the causal situations (`worlds') that are consistent with the listener's knowledge. We will suppose that this listener has asked (perhaps explicitly) the question ``why $\textsf{FACT}$?'', where $\mathsf{FACT}$ is something the listener knows (that is, $\mathsf{FACT}$ is true in all causal situations in $\mathcal{K}$).

\begin{figure}[ht]
    \centering
    \subfloat[$\mathbf{u}_{C}$]{\begin{tikzpicture}
    \tikzset{node style/.style={circle, draw, minimum size=.8cm, fill opacity=1.0}}
    
    \node[node style, fill=red!20] (A) at (0,0) {A};
    \node[node style, fill=red!20] (B) at (1, .75) {B};
    \node[node style, fill=green!20] (C) at (2, 0) {C}; 
    \node[node style, fill=yellow!20] (E) at (1,-1.5) {E};
    
    \draw[->,thick] (A) -- (E);
    \draw[->,thick] (B) -- (E);
    \draw[->,thick] (C) -- (E); 
\end{tikzpicture}}
    \hfill 
        \subfloat[$\mathbf{u}_{A,C}$]
        {\begin{tikzpicture}
    \tikzset{node style/.style={circle, draw, minimum size=.8cm, fill opacity=1.0}}
    
    \node[node style, fill=green!20] (A) at (0,0) {A};
    \node[node style, fill=red!20] (B) at (1, .75) {B};
    \node[node style, fill=green!20] (C) at (2, 0) {C}; 
    \node[node style, fill=yellow!20] (E) at (1,-1.5) {E};
    
    \draw[->,thick] (A) -- (E);
    \draw[->,thick] (B) -- (E);
    \draw[->,thick] (C) -- (E); 
\end{tikzpicture}}
\hfill
        \subfloat[$\mathbf{u}_{B,C}$]{\begin{tikzpicture}
    \tikzset{node style/.style={circle, draw, minimum size=.8cm, fill opacity=1.0}}
    
    \node[node style, fill=red!20] (A) at (0,0) {A};
    \node[node style, fill=green!20] (B) at (1, .75) {B};
    \node[node style, fill=green!20] (C) at (2, 0) {C}; 
    \node[node style, fill=yellow!20] (E) at (1,-1.5) {E};
    
    \draw[->,thick] (A) -- (E);
    \draw[->,thick] (B) -- (E);
    \draw[->,thick] (C) -- (E); 
\end{tikzpicture}}
\hfill
        \subfloat[$\mathbf{u}_{A,B}$]{\begin{tikzpicture}
    \tikzset{node style/.style={circle, draw, minimum size=.8cm, fill opacity=1.0}}
    
    \node[node style, fill=green!20] (A) at (0,0) {A};
    \node[node style, fill=green!20] (B) at (1, .75) {B};
    \node[node style, fill=red!20] (C) at (2, 0) {C}; 
    \node[node style, fill=yellow!20] (E) at (1,-1.5) {E};
    
    \draw[->,thick] (A) -- (E);
    \draw[->,thick] (B) -- (E);
    \draw[->,thick] (C) -- (E); 
\end{tikzpicture}}
\hfill
        \subfloat[$\mathbf{u}_{A,B,C}$]{\begin{tikzpicture}
    \tikzset{node style/.style={circle, draw, minimum size=.8cm, fill opacity=1.0}}
    
    \node[node style, fill=green!20] (A) at (0,0) {A};
    \node[node style, fill=green!20] (B) at (1, .75) {B};
    \node[node style, fill=green!20] (C) at (2, 0) {C}; 
    \node[node style, fill=yellow!20] (E) at (1,-1.5) {E};
    
    \draw[->,thick] (A) -- (E);
    \draw[->,thick] (B) -- (E);
    \draw[->,thick] (C) -- (E); 
\end{tikzpicture}}
    \caption{Bob's epistemic state $\mathcal{K}$ in Example \ref{example: holiday tan}. Red nodes indicate that the variable is `false' (has value $0$), and green `true'. The explanandum (known to have value $1$) is in yellow.}
    \label{fig:knowledge_state_holiday}
\end{figure}
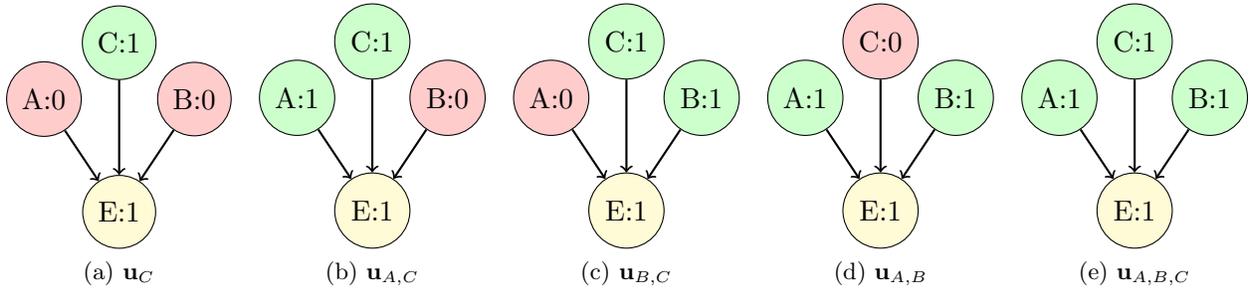

Concretely, consider \autoref{example: holiday tan} once again. Bob knows that Alice is tanned (that $E=1$); let's suppose that he also knows the causal structure is given by $\mathcal{M}_{(A\land B)\lor C}$ (\autoref{fig:holidaytanstructure}). Then his epistemic state $\mathcal{K}$ consists of five pairs $(\mathcal{M}_{(A\land B)\lor C}, -)$:  he is uncertain only about the five possible settings of exogenous variables that would lead to $E=1$,\footnote{These are $\mathbf{u}_{C}=\langle U_{A}=0,  U_{B}=0,  U_{C}=1\rangle$;  $\mathbf{u}_{A,B}=\langle U_{A}=1,  U_{B}=1,  U_{C}=0 \rangle$; $\mathbf{u}_{A,C}=\langle U_{A}=1,  U_{B}=0,  U_{C}=1 \rangle$; $\mathbf{u}_{B,C}=\langle U_{A}=0,  U_{B}=1,  U_{C}=1 \rangle$; and $\mathbf{u}_{A,B,C}=\langle U_{A}=1,  U_{B}=1,  U_{C}=1 \rangle$.} depicted in \autoref{fig:knowledge_state_holiday} (exogenous variables omitted for readability). 

Moving beyond the setup from \citet{halpern2005causesExplanationsCause}, we assume that a second agent -- the \emph{speaker} -- is involved.\footnote{To disambiguate the speaker and listener, we use female pronouns for the speaker and male pronouns for the listener. In examples, we refer to the speaker as `Alice' and the listener as `Bob'.} The phenomenon of explanation is to be elucidated via the interaction between the listener and the speaker. 

To model this interaction we draw upon a body of work at the intersection of linguistics and psychology intended to capture the pragmatics of ordinary conversation. This work, sometimes referred to as the \emph{rational speech acts} (or RSA) framework, can be seen as a formalisation of familiar Gricean ideas from the philosophy of language \citep{frank2012predicting, goodman2016pragmati, degen2023rational}. It involves a hierarchy of conversational agent models, typically beginning with a ``literal'' listener, who simply interprets messages according to their semantic content, before then ascending to a pragmatic speaker, who selects a message taking into account how it will be interpreted by the literal listener. Last comes a pragmatic listener who interprets the pragmatic speaker's utterance as a ``rational speech act''. We follow this presentation here. 

\subsection{The Literal Listener}

Imagine a listener who has posed the question, ``why $\mathsf{FACT}$?'', and now receives a response $m$. Formally, the listener has uncertainty about what the world is like, represented by a probability function $\mathsf{Prior}$ on an epistemic state $\mathcal{K}$ consisting of pairs $(\mathcal{M},\mathbf{u})$. In general, $m$ will be true at some pairs $(\mathcal{M},\mathbf{u})$ (written $\mathcal{M}, \mathbf{u} \models m$) and false at others. What might the listener do with response $m$?

Following the RSA framework, we suppose that there is first a `literal' listener $L0$, who updates his uncertainty toward a ``posterior'' distribution $P_{L0}$ as:\footnote{Here, $\mathbf{1}$ is an ``indicator function,'' equal to $1$ if the expression in the subscript is true, and $0$ otherwise. Meanwhile, `$\propto$' signifies ``is proportional to.'' That is, $P_{L0}(\mathcal{M}, \mathbf{u} \:|\: m) = (1/Z)\cdot \mathsf{Prior}(\mathcal{M},\mathbf{u}) \cdot \mathbf{1}_{\mathcal{M}, \mathbf{u} \models m}$, where $Z$ is a normalising constant.} 
\begin{eqnarray}
\label{eq:literallisten}
    P_{L0}(\mathcal{M}, \mathbf{u} \:|\: m) & \propto & \mathsf{Prior}(\mathcal{M},\mathbf{u}) \cdot \mathbf{1}_{\mathcal{M}, \mathbf{u} \models m} 
\end{eqnarray}
In words, message $m$ leads $L0$ to assign probability $1$ to $m$ being true, and to redistribute probability mass among the remaining possibilities in proportion to his prior beliefs.

To model the interventionist idea that explanations communicate causal information, we constrain the semantic value $\semantics{m}$ of answers to the question ``why $\mathsf{FACT}$?''.\footnote{Formally, $\semantics{m} = \{ (\mathcal{M}, \mathbf{u}) \in \mathcal{K}\:|\: \mathcal{M}, \mathbf{u} \models m\}$.}
Specifically, suppose that $m:=\text{``because $\mathbf{X}=\mathbf{x}$''}$.\footnote{Later in the paper, we will see examples of more complex messages $m$.} We suggest that
\begin{eqnarray}
\label{eq:semantics}
    \semantics{\text{``$\mathsf{FACT}$ because $\mathbf{X}=\mathbf{x}$''}} &= & \{ (\mathcal{M}, \mathbf{u}) \in \mathcal{K}\:|\: \mathcal{M}, \mathbf{u} \models \mathsf{Cause} (\mathbf{X}=\mathbf{x},\mathsf{FACT})\} 
\end{eqnarray}
where $\mathsf{Cause} (\mathbf{X}=\mathbf{x},\mathsf{FACT})$ is true at $(\mathcal{M}, \mathbf{u})$ iff $\mathbf{X}=\mathbf{x}$ is an actual cause of $\mathsf{FACT}$ in $(\mathcal{M}, \mathbf{u})$. As noted above, for the purposes of this paper, virtually any account of actual cause from the literature will be suitable here. 

To illustrate how this works, suppose that Alice says, ``because I was in Rome!'' in answer to Bob's question ``why are you tanned?'' (i.e. ``$E=1$ because $A=1$''). What can Bob conclude? In only two situations -- namely, (d) and (e) in \autoref{fig:knowledge_state_holiday} -- would $A=1$ be an actual cause of $E=1$ (written as $\mathsf{Cause}(A=1,E=1)$). This is because (according to all extant treatments of actual causation) an actual cause needs to be true, and suitable enabling conditions (in this case, $B=1$, that it is sunny in Rome) also need to be true. Thus, Bob reassigns all probability mass to these two possibilities, with their relative probability unchanged from the prior.\footnote{One might expect that a listener like Bob could draw even stronger inferences from Alice's statement, e.g., that possibility (d) is significantly more likely than (e). Such an inference is not logically implied by what is said -- and so has sometimes been labeled an \emph{illusory} disjunctive inference \citep{johnson1999illusory} -- but under reasonable assumptions about conversational dynamics it can be perfectly justifiable \citep{SableMeyer}. There is perhaps some sense in which $A=1$ is a ``better'' cause to cite when $C=0$ than when $C=1$. This type of inference, however, may depend on Bob's ability to think about why Alice said what she did (as opposed to various alternatives she could have said). So it will depend on going beyond the literal meaning that we proposed above, viz. an egalitarian notion of actual causation. Our full model could be used to account for this sort of inference (as hinted at in \autoref{section: abnormal cause selection in conjunctive structures}).}

Even at this relatively simple stage, a literal listener will be able to draw additional inferences due to the causal content of Alice's utterance. In the example above, because $A=1$ must be an actual cause, Bob can infer that $B=1$ must also hold, just as highlighted in \autoref{section: communicative context}. To go beyond what is implied by mere actual causation we will incorporate some degree of higher-order reasoning, which first involves making explicit Alice's predicament as a speaker. 

\subsection{The Speaker}
\label{section: the speaker}

Imagine the speaker $S$ is aware of the listener, and of how he will perform `literal' updates of his beliefs following various possible responses she could give to his ``why?'' question. How should the speaker choose a response?

\subsubsection{Useful Utterances}\label{section: useful utterances}

While much work in the RSA framework has focused on `pure information exchange', a natural thought -- explored in recent work by \cite{sumers2023reconciling} -- is that a cooperative speaker will convey information that helps the listener achieve his goals. Following \citeauthor{sumers2023reconciling}, let us suppose that the listener possesses a decision problem, characterised by a pair $(\mathcal{A},\mathcal{R})$, where $\mathcal{A}$ is a set of actions and $\mathcal{R}:\mathcal{A}\times\mathcal{K}\rightarrow \mathbb{R}$ is a \emph{reward function}, specifying how good some action $a$ is in each possible causal situation. The listener chooses an action $a$ and receives a scalar reward $\mathcal{R}(a,\mathcal{M},\mathbf{u})$, where $(\mathcal{M},\mathbf{u})$ is the actual causal situation. So if he knew which possibility in $\mathcal{K}$ in fact obtained, then he would simply choose the action that returns the highest reward in that situation.

A cooperative speaker will produce an utterance that leads the listener toward better actions, via updating his beliefs. Recall from the previous section that a message $m$ will lead $L0$ to update $\mathsf{Prior}$ to a posterior distribution over $\mathcal{K}$, namely $P_{L0}(-\:|\: m)$. We suppose that the listener will then (approximately) maximise expected reward with respect to $P_{L0}$; the probability with which $L0$ chooses action $a$ given message $m$ is 
\begin{eqnarray}
\label{eq:policy}
    \pi_{L0}(a\:|\: m) &\propto & \exp\bigg(\beta_{L}\cdot \bigg[\sum_{(\mathcal{M}, \mathbf{u}) \in \mathcal{K}} P_{L0}(\mathcal{M}, \mathbf{u} \:|\: m) \cdot R(a,\mathcal{M}, \mathbf{u})\bigg]\bigg), 
\end{eqnarray}
where $\exp(z) = e^z$ is the exponential function, and $\beta_L \in \mathbb{R}^{\geq 0}$ is a ``rationality parameter'', measuring how close the agent is to maximizing expected utility; $L0$ comes ever closer as $\beta_L$ tends to infinity. This choice follows a large body of work in psychology,  economics, and computer science modelling agents as approximate expected utility maximisers (see, e.g., \cite{Luce}).

If $L0$ is employing decision policy $\pi_{L0}$, then the reward that $S$ can expect of $L0$ is given by
\begin{eqnarray}
    U_{S}(m, \mathcal{M}, \mathbf{u}) &= & \sum_{a \in \mathcal{A}}\pi_{L0}(a\:|\: m) \cdot R(a,\mathcal{M}, \mathbf{u}), 
\end{eqnarray}
$S$ ought to choose her utterance $m$ -- in our setting, her answer to the listener's ``why?'' question -- in a way that (approximately) maximises this anticipated utility $U_S$: 
\begin{eqnarray}
    P_{S}(m \:|\: \mathcal{M}, \mathbf{u}) &\propto & \exp \Big( \beta_{S} \cdot U_{S}(m, \mathcal{M}, \mathbf{u})\Big).
\end{eqnarray}
Like in \autoref{eq:policy}, the parameter $\beta_S$ here measures how close $S$ is to optimizing. Most often in this paper, $\beta_L$ and $\beta_S$ will be set to $\infty$, meaning that agents are maximizing.

In the examples we discuss in this paper, we'll assume that the speaker has perfect knowledge of the actual causal situation. We'll also assume that the speaker is perfectly cooperative; that is, she communicates only to be useful to the listener, rather than in pursuit of her own ends. Both assumptions could be dropped in future work.

We close this section with a crucial caveat: talk of `decision problems' and `usefulness' should \textit{not} be taken to imply that our model only applies in cases in which the listener has some concrete real-world choice to make. It's important to emphasise that these decision problems can be very abstract. See, for example, manipulation games -- defined in \autoref{section: selecting decision problems} -- and the discussion in \autoref{section: forward-looking decision problems} and \autoref{section: discussion: Explanations Identify Good Points of Intervention}.

\subsubsection{Production and Processing Costs}
\label{section: production and processing costs}

Part of what is interesting about explanations is that they must be constructed and interpreted by resource-limited agents \citep{Wimsatt,ylikoski2010DissectingExplanatoryPower,potochnik2017IdealizationAimsScience}. One can imagine several ways of incorporating such resource limitations. For instance, $S$ may be aware of the possibility that $L0$ could misinterpret a message, or indeed that $S$ herself may err in her production of the message. In this vein, we could incorporate assumptions $S$ could make about the \emph{channel capacity} for messages from $S$ to $L0$ (see, e.g., \citealt{gibson2019efficiency}).

An alternative, which we will adopt here for the sake of simplicity, is that both production and processing of a message $m$ come with some measurable cost. We put these two sources of cost together into a single function $\mathsf{Cost}$, assigning a scalar $\mathsf{Cost}(m)$ to each possible message $m$. Thereby incorporating costs, the speaker probabilities $P_{S}(m\:|\: \mathcal{M}, \mathbf{u})$ now become
\begin{eqnarray}
    P_{S}(m\:|\: \mathcal{M}, \mathbf{u}) &\propto& \exp \bigg( \beta_{S} \cdot \bigg[U_{S}(m\:|\: \mathcal{M}, \mathbf{u}) - \mathsf{Cost}(m)\bigg]\bigg).
\end{eqnarray}
Our model thus assumes that the speaker and listener will produce and interpret each message correctly; they may just suffer cost in doing so, due to length, obscurity, complexity, etc.\footnote{The cost could also include politeness \citep{yoon2020polite, Chandra2024}.} This assumption could be dropped.

\subsection{What is the Listener's Decision Problem?}
\label{section: selecting decision problems}

The idea that explanation can depend on interests and goals of a listener is a perennial theme across pragmatic approaches  \citep{vanfraassen1980ScientificImage, potochnik2015CausalPatternsAdequate, potochnik2016ScientificExplanationPutting, potochnik2017IdealizationAimsScience,deRegt2017,LombrozoLiquin}. As we aim for a precise framework, we will need to be somewhat concrete about what sorts of decision problems $(\mathcal{A},\mathcal{R})$ agents might face.

In some cases, context will render it common knowledge that the listener has asked a ``why?'' question so as to inform a particular future choice. (We discuss an example like this in \autoref{section: an example: roof replacement}.) Of course, most cases in which a listener asks a ``why?'' question will not be like this -- otherwise the listener would just cut out the explanatory middle-man and ask directly for a recommendation for action!

In some cases, the speaker may be uncertain about which of a set of possible decision problems the listener might face. Thus, we could also imagine that the decision problem $(\mathcal{A},\mathcal{R})$ decomposes into a collection of decision problems $(\mathcal{A}_i,\mathcal{R}_i)$, each with a weight $w_i$, representing how likely it is the listener will face this problem. An action $a$ is now a vector $(\dots, a_i \dots)$, specifying choices for all the decision problems, and the total reward is given by the sum $\mathcal{R}(a,\mathcal{M},\mathbf{u}) = \sum_i w_i \mathcal{R}_i(a_i,\mathcal{M},\mathbf{u})$. (We discuss an example naturally modelled this way in \autoref{section: discussion: Explanations Identify Good Points of Intervention}.)

In most cases, though, the speaker will not be able to identify a specific list of decision problems (e.g. cases in which the speaker has very little knowledge about the listener, or cases in which the explanandum is more abstract -- many cases of scientific explanation are like this). Instead, she might want to impart causal information that broadly promises to be useful. A core interventionist intuition is that capacities for \emph{manipulation} and \emph{control} are of central importance \citep{woodward2003MakingThingsHappen, kirfel2024do,Damico}. We formalise what it might mean to have these capacities with what we call a ``manipulation game''.

In the simplest case, we imagine the listener having asked, ``why $\mathsf{FACT}$?'' suggests that causal information about $\mathsf{FACT}$ is somehow relevant to the decision problems he faces. As a general proxy for whatever those decision problems might be, imagine the following game. The listener is presented with some alternative way the world might have been (that is, different assignments to the exogenous variables in his knowledge state), with probability proportional to their prior likelihood. For each such possibility, the listener must choose some endogenous variable to intervene upon, with the aim of changing the truth value of $\mathsf{FACT}$. In each such case, the listener wins a point just in case he successfully manipulates $\mathsf{FACT}$ in that situation. More formally: 

\begin{definition}[Manipulation Game]
\label{def: simple manipulation game}
A manipulation game is a decision problem $(\mathcal{A},\mathcal{R})$, where: \vspace{-.05in}
\begin{itemize} 
    \item $\mathcal{A}$ is the set of all endogenous variables other than those appearing in $\mathsf{FACT}$.
    \item $\mathcal{R}:\mathcal{A}\times \mathcal{K}\rightarrow \mathbb{R}$ is defined, for a given endogenous variable $X \in \mathcal{A}$ and causal situation $(\mathcal{M},\mathbf{u})\in \mathcal{K}$, as
    \begin{equation*}
        \mathcal{R}(X,\mathcal{M},\mathbf{u})= \sum_{\mathbf{u'} \in \mathsf{Val}(\mathbf{U})} P(\mathbf{u'}) \cdot \mathsf{Manipulates}(X, \mathsf{FACT} \:|\: \mathcal{M}, \mathbf{u'})
    \end{equation*}
    where $\mathsf{Manipulates}(X, \mathsf{FACT} \:|\: \mathcal{M}, \mathbf{u'})$ is a binary-valued function that 
    takes $1$ iff there exists $x \in \mathsf{Val}(X)$ such that either $\mathcal{M}, \mathbf{u'} \models \mathsf{FACT}\land [X=x]\neg\mathsf{FACT}$, 
    or $\mathcal{M}, \mathbf{u'} \models \neg\mathsf{FACT}\land [X=x]\mathsf{FACT}$. That is, $\mathsf{Manipulates}(X, \mathsf{FACT} \:|\: \mathcal{M}, \mathbf{u'})$ takes value $1$ iff there is some intervention to $X$ that changes the value of $\mathsf{FACT}$ in world $(\mathcal{M}, \mathbf{u'})$, and value $0$ otherwise.
\end{itemize}
\end{definition}
\noindent Less formally, the listener submits an endogenous variable to intervene on. For each possible context, he receives a score according to whether manipulating the value of the variable manipulates the value of $\mathsf{FACT}$, weighted by the probability of the context. Note in particular that $\mathcal{R}(X, \mathcal{M}, \mathbf{u})$ is insensitive to the actual context $\mathbf{u}$.\footnote{In future work, the manipulation game could be straightforwardly generalised to model an ability to answer what-if-things-had-been-different questions.}

This reward function is closely related to several models of \emph{causal strength} in the psychological literature. When all the variables are binary it coincides with (a causal version) of the so-called $\Delta P$ measure \citep{WardJenkins,ChengNovick}. Replacing $P(\mathbf{u'})$ with the sampling procedure in \cite{lucas2015ImprovedProbabilisticAccount} (which is sensitive to the actual context $\mathbf{u}$) gives the so-called \emph{counterfactual effect size model} of \cite{quillien2023CounterfactualsLogicCausala}.

\subsection{The Pragmatic Listener and the Goodness of an Explanation}
\label{section:goodness}

Having described the `literal' listener, the speaker, and the range of possible decision problems the speaker could entertain for the listener, we are now ready to present the final component, the `pragmatic' listener. This last agent-type will update his beliefs not directly based on the content of the utterance, but based on the fact that the speaker chose this particular utterance, knowing what she does about the (literal) listener and the decision problems confronting him. 

This pragmatic listener -- who we call $L$ -- updates his prior in a way that is closely analogous to $L0$, the literal listener:
\begin{eqnarray}
P_{L}(\mathcal{M},\mathbf{u} \:|\: m) \propto \mathsf{Prior}(\mathcal{M},\mathbf{u}) \cdot P_{S}(m \:|\: \mathcal{M},\mathbf{u})
\end{eqnarray}
The only difference between $P_{L}$ and $P_{L0}$ is in the second term, with the indicator function on the truth of $m$ (in a world) replaced by the probability that $S$ would utter $m$. This in turn supplies us with an updated agent policy $\pi_L(a\:|\:m)$, identical to the expression for $\pi_{L0}$, except with $P_L$ in place of $P_{L0}$.

While one could ascend further in this theory of mind hierarchy, the second level of pragmatic reasoning -- thinking about a speaker thinking about a literal listener -- has attained privileged status in the empirical and modelling literature on linguistic pragmatics \citep{goodman2016pragmati}. We therefore take it as a reasonable conjecture about how high a typical listener might go in their interpretation of a speaker. 

With this much we are now ready to offer our proposal about explanatory goodness. We submit that this can be measured as: 
\begin{eqnarray}
    \mathsf{Goodness}(m, \mathcal{M}, \mathbf{u}) &=& \sum_{a \in \mathcal{A}}\pi_{L}(a\:|\: m) \cdot R(a,\mathcal{M}, \mathbf{u}) - \sum_{a \in \mathcal{A}}\pi_{\mathsf{Prior}}(a\:|\: m) \cdot R(a,\mathcal{M}, \mathbf{u})
\end{eqnarray}
The first term represents the listener's expected utility in the decision problem(s) he faces after hearing the explanation $m$ and updating his prior distribution $\mathsf{Prior}$ to a distribution $P_{L}(-\:|\:m)$. The second term is a baseline; it represents the listener's expected utility if he hadn't received any explanation, and instead used his prior distribution to decide how to act. So $\mathsf{Goodness}(m, \mathcal{M}, \mathbf{u}) > 0$ iff the explanation was helpful to the listener. When $\mathsf{Goodness}(m, \mathcal{M}, \mathbf{u}) = 0$, the explanation hasn't affected the listener's expected utility. When $\mathsf{Goodness}(m, \mathcal{M}, \mathbf{u}) < 0$, it would have been better for the speaker not to give an explanation at all.

This means that $m$ is a good explanation to the extent that it helps the listener achieve his goals.\footnote{Again, it's important to emphasise that the formalism allows for these goals to be very abstract (e.g. a manipulation game).} So while actual causation plays a fundamental role, it does so only instrumentally, in the way it facilitates information transfer and, ultimately, success at a range of downstream tasks.

As we see it, a definition of explanatory goodness should account for our judgements about the goodness of particular explanations; these judgements comprise the data which philosophers of explanation must answer to. Standard approaches to explanatory goodness identify general rules in this data (``explanatory virtues'') which guide our judgements (better explanations cite unknown information, are more minimal, sit at the right level of description, and so on); an explanation is good to the extent that it accords with these rules. Different philosophers disagree on which rules we should consider -- and how much we should weigh different rules -- when assessing the goodness of an explanation. By contrast, we propose a single, pragmatic definition of explanatory goodness (does the speaker's explanation result in a cognitive update to the listener which helps him to achieve his downstream goals?), and argue in \autoref{section: explaining explanation} that many familiar explanatory virtues emerge from this definition. In certain explanatory settings (namely those in which a listener has a ``weird'' prior or downstream interests), this means that the best explanation according to our account might lack some of the standard explanatory virtues.

\subsection{Summing Up}

To a large extent, the framework we have presented in this section is a 
variation of the rational speech act model, as extended by \cite{sumers2023reconciling}; this very general framework has enjoyed success in modelling many phenomena in human conversational dynamics \citep{degen2023rational}.

What is special about our explanatory setting is twofold. First and most fundamentally, we assume that ``why?'' questions are requests for causal information. This assumption is built into the semantics of ``because'' statements (\autoref{eq:semantics}). Second, listeners are often interested in general issues of manipulation and control, an assumption that is formalised by means of the manipulation game (\autoref{def: simple manipulation game}). This operationalises what it is for listeners to seek a more accurate causal understanding of the world.

In sum, the framework can be seen as a natural, formal combination of reasonably general pragmatic reasoning on the one hand, and the interventionist idea that explanation is about asking and answering what-if-things-had-been-different questions on the other. These two ingredients, combined in the simple way we have done here, is enough to account for some of the most striking features of explanation discussed in the literature. 

\section{Explaining Explanation}\label{section: explaining explanation}

How should we assess an account of explanation? A natural approach starts with the observation that explanation, unlike causation, is \textit{inegalitarian}; it is highly selective. We have strong and systematic intuitions about which explanations are better or worse. Indeed, the literature has pinpointed a handful of `explanatory virtues' that determine when an explanation is more or less apt \citep{ylikoski2010DissectingExplanatoryPower,deRegt2017}, or `lovely' \citep{lipton1991ContrastiveExplanationCausal}.\footnote{Philosophers of science have suggested that there is perhaps no privileged weighting of these (and other) good-making features; they plausibly trade off against one another such that promoting one means neglecting another \citep{Kuhn1970, deRegt2017}.} Accordingly, good explanations
\begin{enumerate}
    \item are sensitive to the downstream interests of the receiver of the explanation (\citealt{deRegt2017, potochnik2017IdealizationAimsScience}; discussed in \autoref{section: modelling downstream interests});
    \item are appropriate to the listener's background knowledge (\citealt{gardenfors1980PragmaticApproachExplanations, halpern2005causesExplanationsCause}; discussed in \autoref{section: modelling background knowledge});
    \item identify explanatory relationships which are invariant across background conditions (\citealt{lewis1986CausalExplanation, ylikoski2010DissectingExplanatoryPower, woodward2006SensitiveInsensitiveCausation, woodward2010CausationBiologyStability}; discussed in \autoref{section: modelling invariance});
    \item involve simpler or more `minimal' explanantia (\citealt{salmon1984ScientificExplanationCausal}; discussed in \autoref{section: modelling minimality and simplicity});
    \item sit at the `right' level of abstraction (\citealt{Yablo, strevens2008DepthAccountScientific, woodward2010CausationBiologyStability, woodward2021ExplanatoryAutonomyRole}; discussed in \autoref{section: modelling proportionality}).
\end{enumerate}
This suggests a desideratum for any account of explanation, namely to make sense of why these features are -- or at least appear to be -- explanatory virtues. In this section, we'll show that the model in the previous section is sufficient to generate the explanatory virtues enumerated above. Our goal here is not to offer a full-fledged defence of our account, but rather to paint the view in enough detail so that its relative merits can be assessed and appreciated.

\subsection{Downstream Interests}
\label{section: modelling downstream interests}

In giving explanations, speakers are sensitive to listeners' downstream interests. Ceteris paribus, explanations which are more useful for the listener are better \citep{deRegt2017}. On our picture, this is unsurprising. Beyond communicating causal information, explanations are no different from other acts of communication: interlocutors who consider their audience's needs are simply better communicators.

This subsection proceeds as follows. In \autoref{section: an example: roof replacement}, we model a toy case in which the speaker is certain which decision problem the listener faces. We then use this case as a starting point for discussion of the fruitfulness of our communication-first account. In \autoref{section: forward-looking decision problems} and \autoref{section: discussion: Explanations Identify Good Points of Intervention}, we show that the model concretises two important aspects of the interventionist framework. In \autoref{section: pragmatic theories of explanation}, we show that the model can make sense of examples which historically motivated pragmatic theories of explanation.

\subsubsection{Back to Roof Replacement}
\label{section: an example: roof replacement}

It's illustrative to show how our account models this phenomenon formally, by considering a case in which it is common knowledge which decision problem the listener faces. Consider the following example, adapted from \citet{faye2007PragmaticrhetoricalTheoryExplanation}:
\begin{example}[Roof Replacement]\label{example: roof replacement}
A thatched house catches fire. The house is located in an area that has been experiencing a drought. Bob asks, ``why did the house catch fire?''. Suppose that Bob knows both that the house had a thatched roof and that it hadn't rained recently (Bob knows the state of the world), but doesn't know enough about how fires start to know if either of these factors are the sorts of factors that cause fires (Bob is uncertain about the world's causal structure).
\end{example}
We can model this situation with three binary endogenous variables $R,D,F$, and two binary exogenous variables $U_R,U_D$. $F$ represents whether the house catches fire; $R$ whether its roof is thatched; and $D$ whether there was a recent drought. Bob knows $R,D,F=1$, but is uncertain about the causal structure; he believes it is possible that (i) $F=1$ iff $R=1$ (structure $\mathcal{M}_{R}$), (ii) that $F=1$ iff $D=1$ (structure $\mathcal{M}_{D}$), (iii) that $F=1$ iff $R=1$ \textit{and} $D=1$ (a \textit{conjunctive} structure, $\mathcal{M}_{\wedge}$), or (iv) that $F=1$ iff $R=1$ \textit{or} $D=1$ (a \textit{disjunctive} structure, $\mathcal{M}_{\vee}$).\footnote{Note that we assume that Bob knows that thatched roofs and drought are the kinds of things that wouldn't prevent fires, regardless of whether or not they cause them; that is, we assume that Bob knows that the value of $F$ is an increasing function of the values of $R,D$.} Bob's knowledge state $\mathcal{K}$ is depicted in \autoref{fig:knowledge_state_roof_replacement}, 
with exogenous variables omitted for readability.\footnote{Interestingly, this example helps illustrate the counterintuitiveness of (EX2) in HP's analysis. When (e.g.) the actual causal structure is $\mathcal{M}_{R}$, citing $R=1$ as a cause of $F=1$ seems like the only good explanation. But (EX2) always rules $R=1$ out as an explanation, since $R=1$ is not an actual cause of $F=1$ in another structure, namely $\mathcal{M}_{D}$.} We suppose, for the sake of simplicity, that Bob has a uniform prior over these four causal structures.\footnote{Nothing hinges on this; the arguments below would apply with arbitrary prior distributions with full support.}

\begin{figure}[ht!]
\captionsetup[subfigure]{labelformat=empty}
    \centering
    \subfloat[][R only]{\begin{tikzpicture}
    \tikzset{node style/.style={circle, draw, fill opacity=1.0}}
    
    \node[node style, fill=green!20] (A) at (0,-.5) {R};
    \node[node style, fill=green!20] (B) at (2,-.5) {D};
    \node[node style, fill=yellow!20] (E) at (1,-1.5) {F};
    \node at (1,-.75) {$\mathcal{M}_{R}$};
    
    \draw[->,thick] (A) -- (E);
\end{tikzpicture}}\hspace{.25in}
\subfloat[][D only]{\begin{tikzpicture}
    \tikzset{node style/.style={circle, draw, fill opacity=1.0}}
    
    \node[node style, fill=green!20] (A) at (0,-.5) {R};
    \node[node style, fill=green!20] (B) at (2,-.5) {D};
    \node[node style, fill=yellow!20] (E) at (1,-1.5) {F};
    \node at (1,-.75) {$\mathcal{M}_{D}$};
    
    \draw[->,thick] (B) -- (E);
\end{tikzpicture}}\hspace{.25in}
\vspace{.25in}
        \subfloat[][conjunctive]{\begin{tikzpicture}
    \tikzset{node style/.style={circle, draw, fill opacity=1.0}}
    
    \node[node style, fill=green!20] (A) at (0,-.5) {R};
    \node[node style, fill=green!20] (B) at (2,-.5) {D};
    \node[node style, fill=yellow!20] (E) at (1,-1.5) {F};
    \node at (1,-.75) {$\mathcal{M}_{\wedge}$};
    
    \draw[->,thick] (A) -- (E);
    \draw[->,thick] (B) -- (E);
\end{tikzpicture}}\hspace{.25in}
    \subfloat[][disjunctive]{\begin{tikzpicture}
    \tikzset{node style/.style={circle, draw, fill opacity=1.0}}
    
    \node[node style, fill=green!20] (A) at (0,-.5) {R};
    \node[node style, fill=green!20] (B) at (2,-.5) {D};
    \node[node style, fill=yellow!20] (E) at (1,-1.5) {F};
    \node at (1,-.75) {$\mathcal{M}_{\vee}$};
    
    \draw[->,thick] (A) -- (E);
    \draw[->,thick] (B) -- (E);
\end{tikzpicture}}
\caption{Bob's epistemic state $\mathcal{K}$ in Example \ref{example: roof replacement}. Note that $R=1$ and $D=1$ in all four worlds (i.e. Bob knows the context is $\mathbf{u}_{D,R}$).}
\label{fig:knowledge_state_roof_replacement}
\end{figure}

Suppose that Bob's own roof is thatched, and he's wondering whether to replace it. He lives nearby to the house that caught fire and believes, reasonably, that the same factors which caused the house to catch fire are causally responsible for whether or not his house will catch fire in the future. He would prefer to replace his roof ($a_{\text{replace}}$) if thatched roofs cause fires, but otherwise he'd prefer not to pay the expense ($a_{\text{don't replace}}$). We represent this decision problem with the pay-off matrix shown in \autoref{table: Bob's decision problem in roof replacement}.
\begin{table}[ht!]
\caption{Bob's decision problem in the roof replacement example.}
\label{table: Bob's decision problem in roof replacement}
    \centering
    \begin{tabular}{lrrrr}
\toprule
 & $\mathcal{M}_{R}$ & $\mathcal{M}_{D}$ & $\mathcal{M}_{\wedge}$ & $\mathcal{M}_{\vee}$ \\
\midrule
$a_{\text{replace}}$ & $0$ & $0$ & $0$ & $0$ \\
$a_{\text{don't replace}}$ & $-1$ & $1$ & $-1$ & $-1$ \\
\bottomrule
\end{tabular}
\end{table}

Focus on the case where the actual causal structure is $\mathcal{M_{\wedge}}$ (that is, $F=1$ iff \textit{both} $R=1$ and $D=1$). Suppose that Alice is deciding between two utterances, ``$F=1$ because $R=1$'' and ``$F=1$ because $D=1$'', with the following interpretations:
\begin{align*}
    \semantics{\text{``$F=1$ because $R=1$''}} &= \{\mathcal{M}_{R}, \mathcal{M}_{\wedge}, \mathcal{M}_{\vee}\} \\
    \semantics{\text{``$F=1$ because $D=1$''}} &= \{\mathcal{M}_{D}, \mathcal{M}_{\wedge}, \mathcal{M}_{\vee}\}.
\end{align*}
Note that both of Alice's utterances allow Bob to eliminate equally likely (according to his prior) non-actual worlds from consideration ($\mathcal{M}_{R}$ and $\mathcal{M}_{D}$, respectively). So $P_{L0}(\mathcal{M_{\wedge}} \:|\: R=1) = P_{L0}(\mathcal{M_{\wedge}} \:|\: D=1)$ (i.e. citing either cause is equally \textit{informative} to Bob as to the true causal structure). But we have $U_{S}(R=1, \mathcal{M_{\wedge}}) > U_{S}(D=1, \mathcal{M_{\wedge}})$, as is easily seen; it is more \textit{useful} to Bob to learn that the roof's being thatched was a cause. This is because it allows him to make the sensible decision to replace his roof (formally, we have $\pi_{L0}(a_{\text{replace}} \:|\: R=1) > \pi_{L0}(a_{\text{replace}} \:|\: D=1)$ for any value of $\beta_{L} > 0$).

This example illustrates a general phenomenon: two utterances can be equally \textit{informative} to the listener, but not equally \textit{useful} to him. Eliminating $\mathcal{M}_{D}$ is much more useful to Bob than eliminating $\mathcal{M}_{R}$, because the decision problem he possibly faces is sensitive to whether or not $R=1$ is a cause of $F=1$, but not to whether or not $D=1$ is a cause of $F=1$.\footnote{Which causal facts the decision problem is sensitive to can be seen by looking at the columns of the pay-off matrix. When the decision problem is insensitive to a causal fact (as with the fact that $D=1$ is a cause of $F=1$), the pay-off matrix has the same columns even as the causal fact varies (so we see that the structures $\mathcal{M}_{\wedge}$ and $\mathcal{M}_{R}$ have the same columns). By contrast, when the decision problem is sensitive to a causal fact (as with the fact that $R=1$ is a cause of $F=1$), there are cases in which the columns of the pay-off matrix vary with that causal fact (so we see that the structures $\mathcal{M}_{\wedge}$ and $\mathcal{M}_{D}$ have different columns).}

\subsubsection[Forward-Looking Decision Problems and Invariance]{Forward-Looking Decision Problems and Invariant Type-Level Causal Relationships}\label{section: forward-looking decision problems}

It's worth spelling out in more detail an assumption in the way we model \autoref{example: roof replacement}. Technically speaking, Bob's decision problem is sensitive to type-level causal relationships in a closely related causal scenario: what factors could causally affect whether \textit{his own house} will catch fire. In modelling the decision problem as sensitive to the type-level causal facts concerning the house which actually caught fire, we implicitly assume that these type-level causal relationships are unchanged between that house and Bob's house (and that this is common knowledge between Bob and Alice).

This is because Bob's decision problem is \textit{forward-looking}; it is sensitive to what could happen in the future, under the assumption that any type-level causal relationships will stay the same as in the past. As we see it, this assumption (that many type-level causal relationships will remain unchanged in decision problems the agent might face) underlies and concretises the interventionist suggestion that causal information has practical utility. Asking ``why?'' questions about the past gives you information about type-level causal relationships which will be useful for the future \citep[Ch.1]{woodward2003MakingThingsHappen}; as the example above shows, this can be true even when the explanandum itself is a singular event (rather than a regularity). Interestingly, not all decision problems are forward-looking in this way; many will be \textit{backwards-looking}, especially those that involve the attribution of responsibility (see \autoref{example: late meeting} and \autoref{example: roommate theft}).

\subsubsection{Explanations Identify Good Points of Intervention}
\label{section: discussion: Explanations Identify Good Points of Intervention}

What's going on in the example above is not merely that Alice uses her reply to Bob as a way of communicating a recommendation (that he should replace his roof). Rather, as the formalism makes clear, Bob's decision whether or not to replace his roof proceeds via his update on the causal information Alice provides to him. We can make this more explicit by considering a variant of the case above. Suppose now that Alice is uncertain about the precise decision problem Bob faces. She still knows he is considering whether to replace his roof or not, but considers two situations possible.

The first situation is that described above, and modelled by the pay-off matrix in \autoref{table: Bob's decision problem in roof replacement}. In this case, Bob will replace his roof if it would cause a fire, but otherwise would prefer not to spend the money replacing it. Call this decision problem $(\mathcal{A}^{(1)}, \mathcal{R}^{(1)})$. In the second situation, by contrast, Bob wants his house to burn down (say, he is tired of upkeep effort and wants a big insurance pay-out), but would otherwise prefer to replace his roof (say, for aesthetic reasons). We can model this using a pay-off matrix similar to that in \autoref{table: Bob's decision problem in roof replacement}, but where the pay-offs in the two rows have been flipped. Call this decision problem $(\mathcal{A}^{(2)}, \mathcal{R}^{(2)})$ (note that $\mathcal{A}^{(2)}=\mathcal{A}^{(1)}$).

We can model the case in which Alice is uncertain between these two decision problems in the manner described in \autoref{section: selecting decision problems}; let's suppose that Alice thinks the chance that Bob is facing decision problem $(\mathcal{A}^{(1)}, \mathcal{R}^{(1)})$ is $\frac{1}{2}$, and the chance that he's facing decision problem $(\mathcal{A}^{(2)}, \mathcal{R}^{(2)})$ is $\frac{1}{2}$, such that each individual decision problem's reward function contributes equally to Bob's total reward (nothing hinges on this choice).

Note that when the actual causal structure is $\mathcal{M}_{\wedge}$, the best action for Bob to select is $a_{\text{replace}} \in \mathcal{A}^{(1)}$ (as we saw above) and $a_{\text{don't replace}} \in \mathcal{A}^{(2)}$ (i.e. he should replace his roof when he doesn't want his house to burn down, and fail to replace it when he does). Using identical reasoning to that in the previous section, we can see that -- in this causal structure -- it's better for Alice to cite $R=1$ as a cause of $F=1$ than it is for her to cite $D=1$. The point is that information about the roof is more valuable to Bob than the information about the drought \textit{in both the decision problems he could face}.

This discussion clarifies the interventionist suggestion that explanations often communicate effective points of intervention \citep{woodward2003MakingThingsHappen, kirfel2024do}. The point is not that explanations contain direct recommendations for action, but rather that (ceteris paribus) a better explanation will cite causal facts which the listener's decision problem is sensitive to. This means that what it is good for the listener to do is sensitive to the world's causal structure. The most obvious way this can happen is when the information provided by the speaker pertains to the effects of the listener's actions on the world; we saw this in \autoref{example: roof replacement}, when Alice conveyed to Bob that his decision to replace his roof would have a causal effect on his house's catching fire. So information which is useful to the listener often (but of course not always) relates to the effects of causal interventions available to him.

\subsubsection{Pragmatic Theories of Explanation}\label{section: pragmatic theories of explanation}

Finally, note that varying the decision problem faced by the listener is sufficient to account for many of the examples which historically motivated pragmatic theories of explanation. Consider the following example from \citet{hanson1958PatternsDiscovery}, cited by \citet[p.125]{vanfraassen1980ScientificImage} (\citealt{hilton1990ConversationalProcessesCausal} discusses a similar example):
\begin{quote}
    ``There are as many causes of $x$ as there are explanations of $x$. Consider how the cause of death might have been set out by a physician as `multiple haemorrhage', by the barrister as `negligence on the part of the driver', by a carriage‐builder as `a defect in the brakeblock construction', by a civic planner as `the presence of tall shrubbery at that turning'.''
\end{quote}
Van Fraassen takes this to show that there is no common core to explanation, but rather that the appropriate relation (which he calls ``relevance'') between explanans and explanandum varies between communicative contexts. We offer a simpler diagnosis: the characters above all cite different parts of a single larger causal model in giving their explanations (i.e. the explanatory relation is one of causal dependence in all the cases). Which part of the causal model it is better for the character to cite depends on the decision problem her listener faces, which the reader infers from the character's occupation.\footnote{In addition, one might think that the characters \textit{know} different parts of the causal model, based on their differing expertise. In the examples we discuss in this paper, the speaker has perfect knowledge of the actual causal situation; as discussed in \autoref{section: useful utterances}, the model could be extended to account for cases of partial knowledge.} As readers, we naturally suppose that the physician is preparing an autopsy report which will be used by the police to decide whether or not to open a criminal case; the barrister is speaking to a jury who are deciding whether or not to convict the driver; the carriage builder is speaking to a team of engineers who will design the next model of the carriage; and so on.

Finding out more information about the communicative context the characters find themselves in makes this clearer; if the physician had happened to be a witness to the crash, for example, and was asked to testify as to whether the driver was distracted, it would be unhelpful for her to start talking about the victim's internal injuries, whatever her occupation. In other words, the information about the characters' occupation can be superseded by information about their audience in assessing the goodness of their explanation. What's really at stake is whether the information the character provides is \textit{useful} for those receiving the explanation.

As compared to van Fraassen's account, then, our account treats explanations as acts of communication which cite causal information about the explanandum. We treat the inegalitarian intuition that some explanations are better than others as tracking the communicative act's downstream usefulness to the listener. This circumvents a common critique of van Fraassen's account, namely that he doesn't characterise the relevance relation between explanans and explanandum, except to point to its dependence on contextual factors through the use of a handful of examples. The worry is that this renders it unconstrained, such that anything counts as explaining anything else with a suitably gerrymandered choice of relevance relation \citep{kitcher1987VanFraassenExplanation}. Indeed, it seems like getting clear on this relevance relation is exactly what's at stake in giving a theory of explanation \citep{strevens2008DepthAccountScientific}. In particular, one might hope that a pragmatic theory of explanation would give a systematic account of how explanatory relevance depends on contextual factors.

It's of course open to a proponent of van Fraassen's view to adopt the same strategy as we do here, cashing out relevance in terms of an interventionist account of causation (one that models causal relations using SCMs).

In some sense, this idea (taking van Fraassen's view – or one like it – and combining it with an interventionist account of causation) is exactly the starting point for our account. Note that this only determines whether an act of communication counts as an explanation; it doesn't by itself give any means for comparing multiple competing causal explanations of the explanandum (that is, it doesn't help at all with modelling causal selection). What our account adds is a worked out formal story about not merely what constitutes an explanation, but also how to compare competing explanations (namely by calculating the listener's cognitive update on receiving the explanation and then evaluating the extent to which this update helps him achieve his goals). Not only is this proposal more concrete than any prior pragmatic accounts, it is significantly different from van Fraassen's own account of how to compare competing explanations, as outlined in \citep[Ch.5, Section 4.4]{vanfraassen1980ScientificImage}. There, van Fraassen sketches three criteria for assessing the goodness of an explanation. He doesn't give a means for aggregating these criteria.\footnote{As he puts it himself, ``the account I am able to offer is neither as complete nor as precise as one might wish. Its shortcomings, however, are shared with the other philosophical theories of explanation I know'' (p.151).} Using the notation in this paper, though, we can try to reconstruct these criteria as follows. Suppose we have two competing explanations $m$ and $m'$ of a given explanandum $\mathsf{FACT}$, expressing propositions $\semantics{m}$ and $\semantics{m'}$ respectively.

The first criterion states that $\semantics{m}$ is better than $\semantics{m'}$ if it is higher probability than $\semantics{m'}$, relative to our background knowledge. In our framework, this is equivalent to saying
    \begin{equation*}
        \mathsf{Prior}(\semantics{m}) > \mathsf{Prior}(\semantics{m'})
    \end{equation*}

For the second criterion, van Fraassen invites us to imagine that we didn't know that $\mathsf{FACT}$ held, but that our background knowledge was otherwise unchanged.\footnote{We read him as intending something like belief contraction by $\mathsf{FACT}$ here, though his informal discussion of course precedes standard formal treatments of contraction \citep{Gardenford1988}.} Then $\semantics{m}$ is better than $\semantics{m'}$ if (relative to this new background knowledge) $\semantics{m}$'s being true makes it more likely that the $\mathsf{FACT}$ is true (relative to other members of the contrast class) than $\semantics{m'}$'s being true. Van Fraassen doesn't settle on a formalisation of this condition.

The third criterion states that $\semantics{m}$ is better than $\semantics{m'}$ if $\semantics{m}$ ``screens off'' $\mathsf{FACT}$ from $\semantics{m'}$, but $\semantics{m'}$ doesn't screen off $\mathsf{FACT}$ from $\semantics{m}$. In our framework, this occurs when the following conditions hold:
\begin{align*}
    \mathsf{Prior}(\mathsf{FACT}\:|\:\semantics{m}, \semantics{m'}) &= \mathsf{Prior}(\mathsf{FACT}\:|\:\semantics{m}) \\
    \mathsf{Prior}(\mathsf{FACT}\:|\:\semantics{m}, \semantics{m'}) &\neq \mathsf{Prior}(\mathsf{FACT}\:|\:\semantics{m'})
\end{align*}
It's hard to know how to apply the second criterion (or to give it a causal construal) without greater detail. And the first and third criteria yield strange predictions. The third criterion, for example, has the consequence that we should always prefer explanations which cite later points in a causal chain.\footnote{In fact, some empirical evidence shows the opposite preference \citep{McClure2007Judgments, Hilton2010Selecting, pacer2017ockhams}.} The first criterion implies that we should prefer explanations which cite causes which are more likely to be true according to our prior knowledge, whereas in the case of conjunctive causal structures a growing body of empirical evidence demonstrates the opposite preference (we discuss this in detail in \autoref{section: abnormal cause selection in conjunctive structures}) . Moreover, there are many phenomena van Fraassen's criteria seem totally unable to account for (e.g. the fact that better explanations tend to cite proportional causes, or the fact that better explanations tend to cite unknown causes). Perhaps van Fraassen intends these phenomena to be accounted for by his notion of relevance (e.g. he could stipulate that only causes at the ``right level'', or causes which are unknown to the listener, will be explanatorily relevant to the explanandum); again, though, we run into the issue of what exactly relevance is supposed to be, and how it is determined. So whilst our high-level proposal is very much in the spirit of van Fraassen's account, its details about how to compare explanations are very different; it is these details that matter for evaluating its consequences.

\subsection{Background Knowledge}
\label{section: modelling background knowledge}

Explanations are better if they are appropriate to the listener's background knowledge. In particular, better explanations tend to cite information which is not already known to the listener. As we've seen (\autoref{section: explanation and causation}), existing formal accounts of explanation simply stipulate that explanations involve the provision of unknown information \citep{gardenfors1980PragmaticApproachExplanations,halpern2005CausesExplanationsStructuralModela}. By contrast, our model does not require that speakers only cite unknown information. Instead, this fact emerges from our model of the dynamics of communication, as a general consequence of the fact that speakers aim at usefulness. 

To see this, consider another variant of \autoref{example: roof replacement}. As in \autoref{section: discussion: Explanations Identify Good Points of Intervention}, let's suppose that Alice thinks it possible that Bob is facing one of two decision problems. The first is the same as that in \autoref{section: an example: roof replacement}, specified by the pay-off matrix in \autoref{table: Bob's decision problem in roof replacement}. Denote it $(\mathcal{A}^{(1)}, \mathcal{R}^{(1)})$. For the second, suppose that Bob is considering whether or not to move to an area which is unaffected by drought. He'd rather not move, but would prefer to do so if droughts cause fires. We represent this with the pay-off matrix in \autoref{table: Bob's decision problem in moving}.
\begin{table}[ht!]
\caption{Bob's decision problem in the moving example.}
\label{table: Bob's decision problem in moving}
    \centering
    \begin{tabular}{lrrrr}
\toprule
 & $\mathcal{M}_{R}$ & $\mathcal{M}_{D}$ & $\mathcal{M}_{\wedge}$ & $\mathcal{M}_{\vee}$ \\
\midrule
$a_{\text{move}}$ & $0$ & $0$ & $0$ & $0$ \\
$a_{\text{stay}}$ & $1$ & $-1$ & $-1$ & $-1$ \\
\bottomrule
\end{tabular}
\end{table}
Denote this second problem $(\mathcal{A}^{(2)}, \mathcal{R}^{(2)})$.

Suppose, as in the examples above, that the actual causal structure is $\mathcal{M}_{\wedge}$. Note that the first decision problem  $(\mathcal{A}^{(1)}, \mathcal{R}^{(1)})$ is sensitive only to whether $R=1$ is a cause of $F=1$, whereas the second decision problem $(\mathcal{A}^{(2)}, \mathcal{R}^{(2)})$ is sensitive only to whether $D=1$ is a cause of $F=1$. So unlike the case in \autoref{section: discussion: Explanations Identify Good Points of Intervention}, in which knowledge about whether $R=1$ was a cause of $F=1$ was more useful to Bob in both decision problems, here which cause  would be better for Alice to cite depends on which decision problem Bob actually faces. In particular, when Alice thinks it equally likely that Bob faces the first decision problem as it is that he faces the second, then $U_{S}(R=1, \mathcal{M_{\wedge}}) = U_{S}(D=1, \mathcal{M_{\wedge}})$.

Suppose, though, that Bob learns that drought is a cause of the fire (i.e. that $D=1$ is a cause of $F=1$). We model this by supposing $\mathsf{Prior}$ is instead uniform over $\{\mathcal{M}_{D},\mathcal{M}_{\wedge},\mathcal{M}_{\vee}\}$). In this case, we will have $\mathsf{Goodness}(R=1, \mathcal{M_{\wedge}}) > \mathsf{Goodness}(D=1, \mathcal{M_{\wedge}})$.
In other words, given Bob knows that $D=1$ is a cause of $F=1$, citing $D=1$ no longer improves his performance on $(\mathcal{A}^{(2)},\mathcal{R}^{(2)})$ in $\mathcal{M}_{\wedge}$, whereas citing $R=1$ continues to improve his performance on $(\mathcal{A}^{(1)},\mathcal{R}^{(1)})$.

So our model accounts for the intuition that it is better to cite causes which are unknown to the listener, without simply stipulating this. In fact, as we will see next, there are cases in which a good explanation can cite an event which is known to the listener to be a cause of the explanandum (a counterexample to (EX4) on the HP analysis). It is a virtue of our proposal that it allows for these cases to occur.

\subsubsection{Known causes can be informative}
\label{section: informative known causes}

Consider the following:
\begin{example}[Late Meeting]\label{example: late meeting}
Bob is late to meet Alice and Charlie, and can tell that Charlie is cross at him when he arrives. He knows that his tardiness is a cause of Charlie's crossness (Charlie is famously punctual), but is unsure if it's sufficient by itself. In particular, he's wondering if Charlie's cross that he forgot their birthday the previous week. When Charlie is out of the room, Bob asks Alice, ``why is Charlie cross at me?''
\end{example}
We model this situation with binary variables $T,B,C$ where $T$ represents Bob's tardiness, $B$ represents his having forgotten Charlie's birthday, and $C$ represents Charlie's being cross. Bob knows that $B,C,T=1$, but is uncertain whether the causal structure is given by $\mathcal{M}_{T}$ (his tardiness is necessary and sufficient for Charlie's crossness) or by $\mathcal{M}_{\wedge}$ (both his tardiness and his having forgotten Charlie's birthday are individually necessary, and jointly sufficient). See \autoref{fig:knowledge_state_late_meeting}; suppose that $\mathsf{Prior}(\mathcal{M}_{T}) = \mathsf{Prior}(\mathcal{M}_{\wedge})$, for the sake of simplicity.

\begin{figure}[ht!]
\captionsetup[subfigure]{labelformat=empty}
    \centering
    \subfloat[][T only]{\begin{tikzpicture}
    \tikzset{node style/.style={circle, draw, fill opacity=1.0}}
    
    \node[node style, fill=green!20] (A) at (0,-.5) {T};
    \node[node style, fill=green!20] (B) at (2,-.5) {B};
    \node[node style, fill=yellow!20] (E) at (1,-1.5) {C};
    \node at (1,-.75) {$\mathcal{M}_{T}$};
    
    \draw[->,thick] (A) -- (E);
\end{tikzpicture}}\hspace{.25in}
\subfloat[][conjunctive]{\begin{tikzpicture}
    \tikzset{node style/.style={circle, draw, fill opacity=1.0}}
    
    \node[node style, fill=green!20, fill=green!20] (A) at (0,-.5) {T};
    \node[node style, fill=green!20, fill=green!20] (B) at (2,-.5) {B};
    \node[node style, fill=yellow!20] (E) at (1,-1.5) {C};
    \node at (1,-.75) {$\mathcal{M}_{\wedge}$};
    
    \draw[->,thick] (A) -- (E);
    \draw[->,thick] (B) -- (E);
\end{tikzpicture}}
\caption{Bob's epistemic state $\mathcal{K}$ in Example \ref{example: late meeting}.}
\label{fig:knowledge_state_late_meeting}
\end{figure}
Let's suppose that Alice doesn't have long before Charlie gets back in the room; she has two utterances available to her, ``$C=1$ because $T=1$'' and ``$C=1$ because $B=1$'', with interpretations as follows:
\begin{align*}
    \semantics{\text{``$C=1$ because $T=1$''}} &= \{\mathcal{M}_{T}, \mathcal{M}_{\wedge}\} \\
    \semantics{\text{``$C=1$ because $B=1$''}} &= \{\mathcal{M}_{\wedge}\}.
\end{align*}
When Bob asks ``why is Charlie cross at me?'', what is his decision problem? It seems natural to suppose that he is asking the question to inform his apology. In particular, let's suppose that he is already planning to apologise to Charlie for being late, but is wondering whether to apologise for forgetting Charlie's birthday too.\footnote{Note this is an example of a \textit{backwards-looking} decision problem; it is sensitive to token-level causal relationships involving a past event, rather than type-level relationships.} If the actual causal structure is $\mathcal{M}_{T}$ (that is, if Bob's tardiness is the sole cause of Charlie's crossness), he would rather not remind Charlie that he forgot his birthday. But if the actual causal structure is $\mathcal{M}_{\wedge}$ (that is, Charlie is cross because Bob was late \textit{and} forgot his birthday), Bob would rather apologise for both things. We can represent this in the pay-off matrix in \autoref{table: Bob's decision problem in apology}, where $a_{\text{tardiness}}$ represents Bob's apologising for being late alone, and $a_{\text{both}}$ represents Bob's apologising for forgetting Charlie's birthday in addition.
\begin{table}[ht!]
\caption{Bob's decision problem in the apology example.}
\label{table: Bob's decision problem in apology}
    \centering
    \begin{tabular}{lrr}
\toprule
 & $\mathcal{M}_{T}$ & $\mathcal{M}_{\wedge}$ \\
\midrule
$a_{\text{tardiness}}$ & $1$ & $-1$ \\
$a_{\text{both}}$ & $-1$ & $1$ \\
\bottomrule
\end{tabular}
\end{table}
Suppose the actual world is given by $\mathcal{M}_{T}$. Then intuitively the best explanation is ``because you were late'' ($T=1$). Importantly, the point is not merely that this is the only explanation available to Alice. Rather, providing this explanation seems \textit{better} than providing no explanation. Intuitively, it allows Bob to infer that the actual world is not $\mathcal{M}_{\wedge}$ (because otherwise Alice would have cited his forgetting Charlie's birthday). Let's see how our model delivers this intuition.

Suppose the actual causal structure is given by $\mathcal{M}_{T}$. We have $P_{L0}(\mathcal{M}_{T}\:|\: T=1) = P_{L0}(\mathcal{M}_{\wedge}\:|\: T=1)$, so $U_{S}(T=1 \:|\: \mathcal{M}_{\wedge})=U_{S}(T=1 \:|\: \mathcal{M}_{T})$; citing $T=1$ doesn't improve the literal listener's performance on the decision problem. But note that
\begin{align*}
    0 = P_{L0}(\mathcal{M}_{T}\:|\: B=1) < P_{L0}(\mathcal{M}_{\wedge}\:|\: B=1) = 1.
\end{align*}
So $U_{S}(B=1 \:|\: \mathcal{M}_{T}) < U_{S}(T=1 \:|\: \mathcal{M}_{T})$ and $U_{S}(B=1 \:|\: \mathcal{M}_{\wedge}) > U_{S}(T=1 \:|\: \mathcal{M}_{\wedge})$; this means that the speaker will cite $B=1$ over $T=1$ iff the causal structure is $\mathcal{M}_{\wedge}$. But then we will have that $P_{L}(\mathcal{M}_{T}\:|\: T=1) > P_{L}(\mathcal{M}_{\wedge}\:|\: T=1)$; the pragmatic listener is able to infer from the fact that the speaker didn't cite $B=1$ that the actual causal structure is $\mathcal{M}_{T}$! Crucially, this means that
\begin{align*}
    \mathsf{Goodness}(T=1, \mathcal{M}_{T}) > 0.
\end{align*}
So it is actively helpful for Alice to cite $T=1$ as a cause of $C=1$, even though this was already known to Bob. Our model delivers the correct intuition.

Of course, the example is contrived: in particular, why suppose that time constraints prohibit Alice from explaining that Charlie is angry because \textit{both} $B=1$ and $T=1$? After all, this is the most intuitive explanation when the actual causal structure is given by $\mathcal{M_{\wedge}}$. But note that this constraint, however contrived here, illustrates a real and important phenomenon: no actual explanation can enumerate every factor in an explanandum's causal history \citep{lewis1986CausalExplanation, potochnik2017IdealizationAimsScience}. The choice of which causal information to include (and which to exclude) is, we suggest, precisely what makes one explanation better than another.

\subsection{Explanatory Relationships and Background Conditions}
\label{section: modelling invariance}

Better explanations tend to cite explanatory relationships which persist across different background conditions. \citet{woodward2003MakingThingsHappen} develops this idea in terms of the \textit{invariance} of a causal relationship; a causal relationship between two variable assignments in an SCM is invariant if it continues to hold as the values of other variables change \citep[see also][]{Vasilyeva}. Woodward suggests that, all else being equal, information about a causal relationship will be more useful for manipulation and control to the degree the causal relationship is invariant.

Our model concretises Woodward's claim that invariant causal relationships are more useful for listeners, via the notion of a \textit{manipulation game} (\autoref{def: simple manipulation game}). Recall that a manipulation game is a type of decision problem which (we argued) arises when the speaker is unsure about precisely why the listener has asked ``why $\mathsf{FACT}$?''.

The speaker supposes that the listener is interested in manipulating whether or not $\mathsf{FACT}$ holds. Concretely, the listener must select an endogenous variable in his knowledge state. He is evaluated on the number of background conditions (settings to endogenous variables in his knowledge state) in which intervening on the variable he has selected would change whether or not $\mathsf{FACT}$ holds (i.e. he is evaluated on his ability to manipulate the explanandum, based on the information received).

In this section, we use manipulation games to show that our model neatly accounts for a body of literature in the empirical literature on causal selection \citep{gerstenberg2020physical,icard2017NormalityActualCausal,kominsky2015superseding,HENNE2019157,quillien2023CounterfactualsLogicCausala, kirfel2022InferenceExplanation}. Specifically, we show that our model correctly predicts that
\begin{itemize}
    \item speakers prefer to cite `abnormal' causes in conjunctive causal structures;
    \item speakers prefer to cite `normal' causes in disjunctive causal structures;
    \item given speakers' preferences, listeners can infer the normality of causes from their knowledge of the causal structure (and vice-versa).
\end{itemize}
This highlights another important way in which our account constitutes a bridge between the philosophical and psychological literatures on explanation.

\subsubsection{Causal Selection Interacts with Normality and Structure}
\label{section: abnormal cause selection in conjunctive structures}

Once again, let's adopt the set-up of \autoref{example: roof replacement}, where Bob's knowledge state is as represented in \autoref{fig:knowledge_state_roof_replacement}. Suppose also that Alice is deciding whether to cite $R=1$ or $D=1$ as a cause of $F=1$. There will be two differences between the cases we've discussed so far and the case we'll consider here. First, the examples above didn't rely on any particular specification of the distribution $P(\mathbf{U})$ on the exogenous variables. In this example, by contrast, we'll suppose that for all the causal situations in Bob's epistemic state, we have $0 < P(U_{R}) < P(U_{D})$.\footnote{Note that $U_{R}, U_{D}$ are the exogenous variables on which $R,D$ depend. So $R = 1$ iff $U_{R} = 1$ and $D=1$ iff $U_{D}=1$. We assume that $U_{R},U_{D}$ are independent.} So $R=1$ is relatively (statistically) `abnormal' and $D=1$ is relatively `normal'. Second, suppose that Alice doesn't know the specific decision problems Bob faces. In this situation, the interventionist suggests that she will give him information useful to manipulating whether or not the house catches fire in similar situations. Should she cite $D=1$ or $R=1$?

We can model this using a manipulation game $(\mathcal{A}, \mathcal{R})$,  as in \autoref{def: simple manipulation game}. There will be two variables which Bob can choose between ($R$ and $D$) and four contexts in which intervening on these variables will be evaluated: $\mathbf{u}_{\emptyset}$, $\mathbf{u}_{R}$, $\mathbf{u}_{D}$ and $\mathbf{u}_{R,D}$, where $\mathbf{u}_{\emptyset}$ denotes the case in which $R,D=0$.\footnote{So we have $P(\mathbf{u}_{\emptyset}) = (1-P(U_{R}))(1-P(U_{D}))$, $P(\mathbf{u}_{R}) = P(U_{R})(1-P(U_{D}))$, $P(\mathbf{u}_{D}) = (1-P(U_{R}))P(U_{D})$, $P(\mathbf{u}_{R,D}) = P(U_{R})P(U_{D})$.} The pay-off matrix is given in \autoref{table: pay-offs for manipulation game in roof example}. Here, $a_{R}$ (respectively, $a_{D}$) denotes Bob's choosing to intervene on $R$ (respectively, $D$).
\begin{table}[ht!]
    \centering
    \caption{Pay-offs for the simple manipulation game for \autoref{example: roof replacement}.}
\label{table: pay-offs for manipulation game in roof example}
    \begin{tabular}{lrrrr}
\toprule
 & $\mathcal{M}_{R}$ & $\mathcal{M}_{D}$ & $\mathcal{M}_{\wedge}$ & $\mathcal{M}_{\vee}$ \\
\midrule
$a_{\text{R}}$ & $1$ & $0$ & $P(U_{D})$ & $1-P(U_{D})$ \\
$a_{\text{D}}$ & $0$ & $1$ & $P(U_{R})$ & $1-P(U_{R})$ \\
\bottomrule
\end{tabular}
\end{table}

Note that $\mathcal{R}(a_{R}, \mathcal{M}_{R})=1$, since changing the value of variable $R$ in structure $\mathcal{M}_{R}$ changes the value of $F$ across all contexts. Conversely, $\mathcal{R}(a_{D}, \mathcal{M}_{R})=0$ since changing the value of variable $D$ in structure $\mathcal{M}_{R}$ has no effect on the value of $F$, regardless of the context. Identical reasoning applies to the column for structure $\mathcal{M}_{D}$. To calculate the entries for the column for structure $\mathcal{M}_{\wedge}$, note that -- in this structure -- changing the value of $R$ successfully manipulates $F$ in contexts $\mathbf{u}_{D}$ and $\mathbf{u}_{R,D}$, and changing the value of $D$ successfully manipulates $F$ in contexts $\mathbf{u}_{R}$ and $\mathbf{u}_{R,D}$. Adding together the probabilities of these contexts gives $P(U_{D})$ and $P(U_{R})$,  respectively. To calculate the entries for the column for structure $\mathcal{M}_{\vee}$, note that -- in this structure -- changing the value of $R$ successfully manipulates $F$ in contexts $\mathbf{u}_{\emptyset}$ and $\mathbf{u}_{R}$, and changing the value of $D$ successfully manipulates $F$ in contexts $\mathbf{u}_{\emptyset}$ and $\mathbf{u}_{D}$. Adding together the probabilities of these contexts gives $1-P(U_{D})$ and $1-P(U_{R})$, respectively.

We have $\pi_{L0}(a_{R}\:|\: R=1) > \pi_{L0}(a_{D}\:|\: R=1)$ and $\pi_{L0}(a_{D}\:|\: D=1) > \pi_{L0}(a_{R}\:|\: D=1)$, as is easily seen. Recall that, by assumption, $P(U_{R})<P(U_{D})$. So we have:
\begin{align*}
    U_{S}(R=1, \mathcal{M}_{\wedge}) &>  U_{S}(D=1, \mathcal{M}_{\wedge}) \\
    U_{S}(R=1, \mathcal{M}_{\vee}) &<  U_{S}(D=1, \mathcal{M}_{\vee})
\end{align*}
Informally, when the actual causal structure is $\mathcal{M}_{\wedge}$, the speaker should cite $R=1$; when the actual causal structure is $\mathcal{M}_{\vee}$, the speaker should cite $D=1$.

This accords with -- and perhaps goes some way toward explaining -- a growing body of evidence in the cognitive science literature that speakers select abnormal causes in conjunctive structures, and normal causes in disjunctive structures \citep{kominsky2015superseding, icard2017NormalityActualCausal,HENNE2019157,gerstenberg2020physical,quillien2023CounterfactualsLogicCausala}. We take it to be compatible with the idea that people judge causal strength by simulating different counterfactual possibilities \citep{icard2017NormalityActualCausal, quillien2023CounterfactualsLogicCausala}; on our picture, they are imagining speaking to a listener playing a simple manipulation game.

\subsubsection{Inferences from Normality}

Aside from production, recent work also shows that a listener can infer a great deal from a speaker's choice of which cause to cite (\citealp{kirfel2022InferenceExplanation}; see also \citealp{davis2025inference}). For instance, people can infer whether a structure is disjunctive or conjunctive (when they know which cause is normal) or infer which cause is normal (when they know the causal structure). Again, our model captures this, via the `pragmatic listener' $L$. For example, suppose Alice cites the abnormal cause (the roof's being thatched, $R=1$). Then, since  $P_{S}(R=1, \mathcal{M_{\wedge}}) > P_{S}(R=1, \mathcal{M_{\vee}})$, we have  
\begin{align*}
    P_{L}(\mathcal{M}_{\wedge}\:|\: R=1) &> 
    P_{L}(\mathcal{M}_{\vee}\:|\: R=1).
\end{align*}
In other words, the pragmatic listener is able to infer that the structure is conjunctive rather than disjunctive, since he takes into account the fact that the speaker will prefer to cite abnormal causes when the structure is conjunctive (and normal causes when it is disjunctive).

A variant of the above example can account for listeners' ability to infer normality of the cited cause from knowledge of the causal structure. Suppose Bob is uncertain about whether $R=1$ or $D=1$ is normal, but knows the causal structure is not disjunctive. That is, $\mathsf{Prior}$ is uniform over
\begin{align*}
    \mathcal{K} = \bigg\{\mathcal{M}^{P(U_{D}) > P(U_{R})}, \mathcal{M}^{P(U_{R}) > P(U_{D})} \:|\: \mathcal{M} \in \{\mathcal{M}_{R}, \mathcal{M}_{D}, \mathcal{M}_{\wedge}\}\bigg\}
\end{align*}
where (e.g.) $\mathcal{M}_{\wedge}^{P(U_{D}) > P(U_{R})}$ is the same as $\mathcal{M}_{\wedge}$, but with $P(U_{D}) > P(U_{R})$ (i.e. with $D=1$ as the normal cause). Suppose the actual causal situation is given by $\mathcal{M}_{\wedge}^{P(U_{D}) > P(U_{R})}$. Then since we have 
\begin{align*}
    P_{S}(R=1, \mathcal{M}_{\wedge}^{P(U_{D}) > P(U_{R})}) > P_{S}(R=1, \mathcal{M}_{\wedge}^{P(U_{R}) > P(U_{D})}),
\end{align*}
we will have
\begin{align*}
    P_{L}(\mathcal{M}_{\wedge}^{P(U_{D}) > P(U_{R})} \:|\: R=1) &> 
    P_{L}(\mathcal{M}_{\wedge}^{P(U_{R}) > P(U_{D})}\:|\: R=1).
\end{align*}
So our model recovers the prediction that listeners can infer the normality of causes from their knowledge of the structure.\footnote{Note that patterns like these violate HP's fourth postulate, EX4. While it is often appropriate to cite unknown factors, this cannot be a hard requirement on explanation.}

\subsection{Minimality and Simplicity} 
\label{section: modelling minimality and simplicity}

Good explanations are often minimal (have no redundant information) and simple (easily comprehended by the listener). We model this using a standard modification to RSA, adding a term to the speaker utility to reflect the \textit{cost} of selecting the message.

Philosophers of science often cash out minimality and simplicity in terms of \textit{length} in some description language \citep{Kitcher1981}.\footnote{There's another sense of simplicity sometimes used by philosophers of science, where a scientific theory's simplicity tracks its ontological parsimony. We use `simplicity' in a different way here.} In this section, we take a more practical, cognitive perspective on minimality and simplicity: different acts of communication resulting in the same cognitive update by the listener can differ in the practical or cognitive resources (of both speaker and listener) required to achieve the update. When we move to this more practical perspective, it's clear that length is at best an imperfect heuristic for the relevant sense of minimality and simplicity. Short messages may be obscure, whilst longer messages may be less cumbersome \citep{ylikoski2010DissectingExplanatoryPower, lage2019evaluationhumaninterpretabilityexplanation}.

\subsubsection{Minimality}

Prior work on explanation often assumes that more minimal explanations are better \citep{halpern2005causesExplanationsCause}. But there are clearly situations in which longer explanations are intuitively better than shorter explanations. This is isn't surprising: after all, a good explanation ought to bear \emph{more} rather than less inferential fruit, a point often stressed in the literature on explanatory generality (e.g., \citealt{Putnam1975, potochnik2017IdealizationAimsScience}). Indeed, psychologists of explanation emphasise that people often prefer explanations that cite more causal mechanisms rather than fewer \citep{zemla2017evaluating,zemla2023not,vrantsidis2024inside}. Consider the following example:

\begin{example}[Milk Theft]\label{example: roommate theft}
Bob returns from work to find that his milk carton is emptier than it was when he left that morning. He knows that the culprit(s) must be at least one (or both) of his roommates, Charlie or Dana. He asks Alice (a neutral, lactose-intolerant fourth roommate, let's say), ``why is my milk gone?''.
\end{example}
We could model this situation with binary variables $C,D,M$ where $C=1$ (respectively, $D=1$) represents Charlie (respectively, Dana) drinking the milk and $M=1$ represents the milk carton's being depleted. Bob knows that the causal structure is given by $\mathcal{M}_{\vee}$ (that is, that at least one roommate's drinking the milk would be necessary and sufficient to deplete it), and knows that $M=1$, meaning at least one of $C=1$ and $D=1$. So there are three possible worlds, given by $\mathbf{u}_{C}$, $\mathbf{u}_{D}$ and $\mathbf{u}_{C,D}$. We represent this in \autoref{fig:knowledge_state_roommate_theft}.

\begin{figure}[ht!]
\captionsetup[subfigure]{labelformat=empty}
    \centering
    \subfloat[\(\mathbf{u}_{C}\)]{%
    \begin{tikzpicture}[scale=1]
        \tikzset{node style/.style={circle, draw, minimum size=.8cm, inner sep=0pt}}
        \node[node style, fill=green!20] (C) at (0,-0.5) {C};
        \node[node style, fill=red!20] (D) at (2,-0.5) {D};
        \node[node style, fill=yellow!20] (M) at (1,-1.5) {M};
        \node at (1,-0.75) {};
        \draw[->,thick] (C) -- (M);
        \draw[->,thick] (D) -- (M);
    \end{tikzpicture}}
    \hspace{.25in}
    \subfloat[\(\mathbf{u}_{D}\)]{%
    \begin{tikzpicture}[scale=1]
        \tikzset{node style/.style={circle, draw, minimum size=.8cm, inner sep=0pt}}
        \node[node style, fill=red!20] (C) at (0,-0.5) {C};
        \node[node style, fill=green!20] (D) at (2,-0.5) {D};
        \node[node style, fill=yellow!20] (M) at (1,-1.5) {M};
        \node at (1,-0.75) {};
        \draw[->,thick] (C) -- (M);
        \draw[->,thick] (D) -- (M);
    \end{tikzpicture}}
    \hspace{.25in}
    \subfloat[\(\mathbf{u}_{C,D}\)]{%
    \begin{tikzpicture}[scale=1]
        \tikzset{node style/.style={circle, draw, minimum size=.8cm, inner sep=0pt}}
        \node[node style, fill=green!20] (C) at (0,-0.5) {C};
        \node[node style, fill=green!20] (D) at (2,-0.5) {D};
        \node[node style, fill=yellow!20] (M) at (1,-1.5) {M};
        \node at (1,-0.75) {};
        \draw[->,thick] (C) -- (M);
        \draw[->,thick] (D) -- (M);
    \end{tikzpicture}}
    \caption{Bob's epistemic state in \autoref{example: roommate theft}}
    \label{fig:knowledge_state_roommate_theft}
\end{figure}

Suppose the actual world is given by $\mathbf{u}_{C,D}$, and Alice knows this. Alice knows that Bob wants to speak to the culprit(s) to get them to stop stealing his milk in the future. In this case, it seems appropriate for her to explain $M=1$ by citing both Charlie and Dana's stealing the milk, rather than by $C=1$ or $D=1$ alone.

The discussion above shows that better explanations need not be minimal. But the fact remains that more minimal explanations are often intuitively better. We account for this this using pragmatic factors, such as communicative pressure.

Let's see how this works. We know Bob wants to confront a roommate iff they have taken his milk; we model this with the pay-offs in \autoref{table: Bob's decision problem in roommate theft} (as above, the precise pay-offs are arbitrary, since we are interested in demonstrating a qualitative effect):
\begin{table}[ht!]
    \centering
    \begin{tabular}{lrrr}
\toprule
 & $\mathbf{u}_{C}$ & $\mathbf{u}_{D}$ & $\mathbf{u}_{C,D}$ \\
\midrule
$a_{\text{Charlie}}$ & $1$ & $-1$ & $0$ \\
$a_{\text{Dana}}$ & $-1$ & $1$ & $0$ \\
$a_{\text{both}}$ & $0$ & $0$ & $1$ \\
\bottomrule
\end{tabular}
\label{table: Bob's decision problem in roommate theft}
\caption{Bob's decision problem in the roommate theft example.}
\end{table}

Let's suppose Alice has three utterances available to her, ``$M=1$ because $C=1$'' (``Charlie took the milk''), ``$M=1$ because $C=1$'' (``Dana took the milk'') and ``$M=1$ because $C=1, D=1$'' (``Charlie and Dana each took the milk''), with the following interpretations:
\begin{align*}
    \semantics{\text{``$M=1$ because $C=1$''}} &= \{\mathbf{u}_{C}, \mathbf{u}_{C,D}\} \\
    \semantics{\text{``$M=1$ because $D=1$''}} &= \{\mathbf{u}_{D}, \mathbf{u}_{C,D}\} \\
    \semantics{\text{``$M=1$ because $C=1, D=1$''}} &= \{\mathbf{u}_{C,D}\}
\end{align*}
It is clear that when the costs associated with each message are equal, we will have $\mathsf{Goodness}(C=1, D=1, \mathbf{u}_{C,D}) > \mathsf{Goodness}(C=1, \mathbf{u}_{C,D}), \mathsf{Goodness}(D=1, \mathbf{u}_{C,D})$. When both roommates took milk, it is better for Bob to know this. But suppose, plausibly, that
\begin{align*}
    \mathsf{Cost}(C=1, D=1) &> \mathsf{Cost}(C=1) = \mathsf{Cost}(D=1),
\end{align*}
(because, e.g., the longer message is more onerous to produce). If the difference in costs is sufficiently high, such that
\begin{align*}
    U_{S}(C=1,D=1, \mathbf{u}_{C,D}) - \mathsf{Cost}(C=1, D=1) &< U_{S}(C=1, \mathbf{u}_{C,D}) - \mathsf{Cost}(C=1) \\
    &= U_{S}(C=1, \mathbf{u}_{C,D}) - \mathsf{Cost}(D=1)
\end{align*}
the speaker will prefer the shorter messages (she will be indifferent between them), even though they are less useful.

The discussion above gives us a way of quantifying the amount of \textit{redundancy} in a candidate explanation $m$, by considering the term
\begin{align*}
    U_{S}(m, \mathcal{M}, \mathbf{u}) - \max_{m' \neq m} U_{S}(m, \mathcal{M}, \mathbf{u})
\end{align*}
If this term is positive, then -- assuming all messages have the same cost -- the speaker will select $m$ over the other messages (as $\beta_{S}\rightarrow \infty$, the message will be selected with certainty). The larger the positive value is, the less redundancy the message contains, in the sense that we could increase its production cost relative to the other messages and it would still be selected by the speaker.

To give a concrete example of a message with high redundancy, suppose that $U_{S}$ is specified using the decision problem $(\mathcal{A}, \mathcal{R})$ in \autoref{example: roof replacement}, with the pay-off matrix in \autoref{table: Bob's decision problem in roof replacement}. Suppose Alice has the utterance ``$F=1$ because \textit{both} $R=1$ \textit{and} $D=1$'' available to her, with interpretation $\{\mathcal{M}_{\wedge}\}$. In this situation, we will have 
\begin{align*}
    U_{S}(R=1, D=1, \mathcal{M}_{\wedge}) &= U_{S}(R=1, \mathcal{M}_{\wedge}),
\end{align*}
since the only information relevant to Bob's decision problem is whether $R=1$ is a cause of $F=1$. In this case, then, if
\begin{align*}
    \mathsf{Cost}(R=1, D=1) &> \mathsf{Cost}(R=1),
\end{align*}
Alice will always prefer the shorter message \text{``$R=1$''}, since it is just as useful to Bob as the more informative message.\footnote{In \autoref{example: holiday tan}, we have another case in which a longer message is redundant. When Alice cites being in Rome as a cause of her tan, she doesn't need to specify that it was also sunny; Bob is able to infer this. Of course, if we supposed there was some cost attached to this inference, and incorporated this into our model, then the more exhaustive explanation would be preferred. See, e.g., \citet[p.214]{ylikoski2010DissectingExplanatoryPower} ``since the power of an explanation is always dependent on the range of counterfactual inferences that the explanatory information enables, the kinds of inferences possible for limited cognitive systems such as humans directly affect what can be explained and understood by such cognitive systems''.}

Note that these considerations also interact with the \textit{normality} of the causes being cited. In \autoref{example: roommate theft} above, if (say) $P(U_{D}=1) >> 0.5 >> P(U_{C}=1)$ (Dana usually takes Bob's food and Charlie rarely does, and this is common knowledge between Alice and Bob), such that $\mathsf{Prior}(\mathbf{u}_{D}) > \mathsf{Prior}(\mathbf{u}_{C,D}) > \mathsf{Prior}(\mathbf{u}_{C})$, then we will have a relatively small difference between $U_{S}(C=1, D=1, \mathbf{u}_{C,D})$ and $U_{S}(C=1, \mathbf{u}_{C,D})$. To see this, note that Bob already assumes Dana has taken the milk, and so -- upon learning that Charlie has taken the milk (i.e. that the actual context is \textit{not} $\mathbf{u}_{D}$) -- believes it is more likely that the actual context is $\mathbf{u}_{C,D}$ than $\mathbf{u}_{C}$. This means that there is more redundancy in the message ``$M=1$ because $C=1, D=1$''.

\subsubsection{Simplicity}

Several authors have observed that choice of wording matters in assessing the quality of an explanation. Consider the following, adapted from \cite{kim1999HempelExplanationMetaphysicsa} and discussed by \cite[p. 217]{woodward2003MakingThingsHappen}:
\begin{example}[Event of the Year]
    Assume that the short circuit caused the fire, and that the short circuit was the most noteworthy event of the year. Bob asks Alice, ``why did the fire happen?'' Intuitively, Alice's utterance ``the short circuit caused the fire'' is a better explanation of the fire's starting than the utterance ``the most noteworthy event of the year caused the fire''.
\end{example}
Woodward suggests that this should be accounted for in terms of invariance; the relationship between ``the short circuit'' and ``the fire'' is invariant across nearby interpretations of ``the short circuit'', in a way that it is not between ``the most noteworthy event of the year'' and ``the fire'' (since ``the most noteworthy event of the year'' could easily have referred to a different event that has nothing to do with the fire). But it's clear that this response isn't satisfactory; invariance describes a causal relationship between two \textit{features of the world} (e.g. events, propositions, etc.), rather than between the phrases used to refer to those features. Moreover, we could easily imagine a situation in which Bob knows what a short circuit is and knows the recent short circuit is widely agreed to be ``the most noteworthy event of the year'', but doesn't know that ``short circuit'' refers to a short circuit. In this case, ``the most noteworthy event of the year caused the fire'' would intuitively be a better explanation for Bob than ``the short circuit caused the fire''.

Thinking in terms of pragmatics provides an easier solution. The speaker is deciding between two utterances $m, m' \in \mathfrak{M}$ which refer to the same feature of the world; that is, we have $\semantics{m}=\semantics{m'}$.\footnote{Formally, for all $(\mathcal{M},\mathbf{u})\in \mathcal{K}$: $\mathcal{M},\mathbf{u} \models m \text{ iff } \mathcal{M},\mathbf{u} \models m'$.}
However, the listener's ability to interpret $m$ might be different from his ability to interpret $m'$; so $m$ and $m'$ -- even though they are identical when interpreted -- incur different processing costs for the listener. In particular, it requires more effort for the listener to recognise that ``the most noteworthy event of the year'' refers to the short circuit than it does ``the short circuit''; there is also more chance that he will misunderstand which part of the world the speaker meant to refer to. So although $m$ and $m'$ have the same truth conditions, $\mathsf{Cost}(m)\neq \mathsf{Cost}(m')$.\footnote{As discussed in \autoref{section: production and processing costs}, our core model is compatible with different ways of formalising a `processing cost' to the listener. For example, one natural idea is that more complex messages have a higher probability of being misinterpreted by the listener;  ``the most noteworthy event of the year'' example could be treated in this way.}

\subsection{Proportionality and Levels of Explanation}
\label{section: modelling proportionality}

Consider the following example, adapted from \citet{Yablo}.\footnote{As \citet{woodward2021ExplanatoryAutonomyRole} observes, some take Yablo to be making a point about the metaphysics of causation here. We follow Woodward in interpreting this as a desideratum on causal explanation.}
\begin{example}
    Alice and Bob see a pigeon peck at a scarlet target. Bob asks ``why did the pigeon peck at the target?''. As it happens, the penguin has been trained to peck at a target iff it is some shade of red. Alice is deciding between the following two explanations of the pigeon's pecking.
    \begin{itemize}
        \item[1.] ``The pigeon pecked at the target because it is scarlet.''
        \item[2.] ``The pigeon pecked at the target because it is red.''
    \end{itemize}
\end{example}
Intuitively, the second explanation is better than the first, even though both involve true causal claims. This intuition is referred to as `proportionality'; it is closely related to the idea that better explanations sit at a more appropriate level of abstraction.

We can model this with two variables, $C$ and $P$. $P$ is a binary variable, representing whether or not the pigeon pecks at the target. $C$ is a variable representing the colour of the target. Let's suppose that it has five possible values: ``empty'' ($0$), ``scarlet'' ($s$), ``crimson'' ($c$), ``blue'' ($b$). Bob knows that $C=s$ (the target is scarlet). Suppose that Bob knows that $P$ depends entirely on the value of $C$; he knows that the pigeon won't peck at an empty target (i.e. that $P=0$ when $C=0$), but doesn't know which values of $C$ (apart from $s$) are sufficient for $P=1$. For each $V \subseteq \{c, b\}$, let $\mathcal{M}_{\{s\} \cup V}$ be the causal model specified by
\begin{align*}
    f_{P}(v) = 1 \quad \text{ iff } v \in \{s\} \cup V.
\end{align*}
For example, $\mathcal{M}_{\{c, s\}}$ is the causal structure corresponding to the case in which the pigeon pecks at the target iff it is crimson \textit{or} scarlet. We can then define Bob's knowledge state
\begin{equation*}
    \mathcal{K} = \{\mathcal{M}_{\{s\}\cup V}\}_{V \subseteq \{c, b\}}.
\end{equation*}
So Bob considers four possible causal situations possible: $\mathcal{M}_{\{s\}}$, $\mathcal{M}_{\{c, s\}}$, $\mathcal{M}_{\{b, s\}}$ and $\mathcal{M}_{\{b, c, s\}}$.

In the example above, the actual causal situation is given by $\mathcal{M}_{\{c, s\}}$; the pigeon pecks at any red target, regardless of whether it is crimson or scarlet. Suppose that Bob is facing a decision problem which is sensitive to the exact causal situation; say, he wants to demonstrate the pigeon's pecking abilities, and is wondering what colour targets to purchase (should he buy scarlet only, scarlet and crimson, scarlet and blue, or targets of any of the colours?). We can model this using the pay-off matrix in \autoref{table: Bob's decision problem in pigeon pecking}, where, for example, $a_{\{c,s\}}$ represents the case in which Bob buys both crimson and scarlet targets to use.

\begin{table}[ht!]
\caption{Bob's decision problem in the pigeon pecking example.}
\label{table: Bob's decision problem in pigeon pecking}
    \centering
    \begin{tabular}{lrrrr}
\toprule
 & $\mathcal{M}_{\{s\}}$ & $\mathcal{M}_{\{c, s\}}$ & $\mathcal{M}_{\{b, s\}}$ & $\mathcal{M}_{\{b, c, s\}}$ \\
\midrule
$a_{\{s\}}$ & $1$ & $0$ & $0$ & $0$ \\
$a_{\{c, s\}}$ & $0$ & $1$ & $0$ & $0$ \\
$a_{\{b, s\}}$ & $0$ & $0$ & $1$ & $0$ \\
$a_{\{b, c, s\}}$ & $0$ & $0$ & $0$ & $1$ \\
\bottomrule
\end{tabular}
\end{table}

Suppose Alice has four utterances available, with the following interpretations:
\begin{align*}
    \semantics{\text{``$P=1$ because the target is scarlet''}} &= \bigg\{\mathcal{M}_{\{s\}}, \mathcal{M}_{\{c, s\}}, \mathcal{M}_{\{b, s\}},\mathcal{M}_{\{b, c, s\}}\bigg\} \\ 
    \semantics{\text{``$P=1$ because the target is red}} &= \bigg\{\mathcal{M}_{\{c, s\}},\mathcal{M}_{\{b, c, s\}}\bigg\} \\ 
    \semantics{\text{``$P=1$ because the target is (scarlet or blue)}} &= \bigg\{\mathcal{M}_{\{b, s\}},\mathcal{M}_{\{b, c, s\}}\bigg\} \\ 
    \semantics{\text{``$P=1$ because the target is coloured}} &= \bigg\{\mathcal{M}_{\{b, c, s\}}\bigg\} \\ 
\end{align*}
Suppose, as in the example, the actual causal structure is $\mathcal{M}_{\{c, s\}}$. Then clearly it is better for the speaker to cite that the target is red than that the target is crimson. In particular, the pragmatic listener $L$ will \textit{correctly} infer that the actual structure is \textit{not} $\mathcal{M}_{\{b, c, s\}}$ from the fact that the speaker \textit{didn't} cite the target's being coloured as the cause. Similarly, if the speaker cites the target's being crimson, the pragmatic listener will \textit{incorrectly} infer that the structure is $\mathcal{M}_{\{s\}}$, since otherwise the speaker would have cited some other cause.\footnote{Readers familiar with RSA will note the similarity between this analysis and the way that RSA accounts for (e.g.) scalar implicature.}

Interestingly, the model predicts interactions between these kinds of inferences and the cost of messages. Suppose that
\begin{equation*}
    \mathsf{Cost}(\text{``$P=1$ because the target is (scarlet or blue)''})
\end{equation*}
is very high. (This seems plausible; after all, there is no single word for (scarlet or blue).) Then when the pragmatic listener hears the speaker cite the target's being scarlet as the cause, he can no longer rule out the causal structure $\mathcal{M}_{\{b, s\}}$ (since the cost constraint means that the speaker wouldn't have cited the disjunction (scarlet or blue) even if this were the structure). This seems to us like the right intuition.

\section{Discussion and Future Work}
\label{section: discussion and future work}

We've introduced a formal pragmatics of explanation, and have shown that it can account for several explanatory virtues, as well as empirical patterns of causal selection. On our picture, candidate explanations are evaluated first and foremost for their usefulness \textit{qua} acts of communication (an ``illocutionary evaluation'', as \citet{achinstein1983NatureExplanation} puts it). To model an explanation, one must represent both the listener's knowledge and the set of decision problems he faces. An important facet of the philosophical problem of explanation, then, consists in accounting for the kinds of things explanations can do for us. Put another way: what are the situations in which it is valuable for someone to seek out causal information, by (e.g.) asking a ``why?'' question?

For pedagogical purposes, the examples we've considered in this paper are contrived in various respects. At this stage, it's natural to wonder: how much of the philosophy of explanation might our model be able to account for, with suitable augmentations? There are several avenues which future work might address; we give three below.

First, much philosophical interest in explanation comes from the central role it plays in scientific inquiry. Applying the model to explanation in the sciences would involve, for example, treating a scientific community as a `listener', and attempting to represent its knowledge and downstream aims more explicitly. This strikes us as a fruitful research area for philosophers of science engaging closely with actual scientific practice.

Second, and relatedly, our framework treats explanations as acts of communication. But of course we can speak of individuals seeking explanations, without any communication taking place (e.g. an individual scientist performing an experiment and analysing the results). We propose that an individual seeking an explanation of $\mathsf{FACT}$ is best thought of as trying to acquire a particular sort of ability, namely the ability to answer a ``why $\mathsf{FACT}$?'' question asked by a hypothetical listener (presumably, a listener with similar goals to her). The claim is not that she is motivated to seek an explanation because of a desire to respond to a listener (or that she sees herself as doing this), but rather that what it is to seek an explanation just is attempting to acquire this ability; this is because the contents of explanations just are the sorts of things that can be communicated, on our view. In order to answer a ``why?'' question successfully, a speaker needs to have an accurate causal model of the explanandum, at a level of description which matches the variables that the listener's decision problems are sensitive to. But she also needs to have a further sort of cognitive ability, namely the ability to describe the causal model using language, in a way that is tailored to the listener's goals and background knowledge. This requires not only causal cognition, but also cognition relating to language \citep{beller2024language}; future work could explore what this cognition looks like, using our framework to test when it has occurred.\footnote{Thanks to an anonymous reviewer for raising this issue.}

Third, our model makes the simplifying assumption that all the causal situations in the listener's knowledge state $\mathcal{K}$ share the same finite set of variables; in particular, the listener is not uncertain about which variables there are, merely about how they relate. But it seems clear that explanations are often most useful precisely when they introduce variables hitherto unknown to the listener; a future version of the model could attempt to model variable introduction explicitly (a natural first attempt would involve representing the knowledge state hierarchically).

Fourth, although we tether the account we develop here to causation, it's clear that the model's formalism can accommodate any dependence describable by an SCM.\footnote{This point is often made about interventionist accounts of explanation \citep{Woodward2018-WOOSVO-2}.} Future work could apply the model to non-causal domains, such as mathematical explanation.

\clearpage

\section*{Acknowledgements}

Thanks to Lio Wong, Ced Zhang, Kartik Chandra and Josh Tennenbaum for helpful discussion. JH presented an early version of the paper at the Stanford CSLI workshop in 2024; thanks to the audience there for useful feedback. TG was supported by grants from the Stanford Institute for Human-Centered Artificial Intelligence (HAI), Toyota Research Institute (TRI), and the Cooperative AI Foundation.

\bibliographystyle{plainnat}
\bibliography{refs}

\end{document}